\def\sec\ond{{\rm s}}

\def\Mpc{{\rm Mpc}}

\def\hMpc{\,h^{-1}\Mpc}
\def\etal{{\frenchspacing\it et al.}}
\def\ie{{\frenchspacing\it i.e.}}
\def\eg{{\frenchspacing\it e.g.}}

\def\rms{{\frenchspacing r.m.s.}}
\def\s{{\small}}
\documentclass[11pt,emulateapj,apj]{emulateapj}
\slugcomment{Not to appear in Nonlearned J., 45.}

\shorttitle{Topology Analysis of the Sloan Digital Sky Survey}
\shortauthors{Park {\etal}}

\begin{document}
%\twocolumn[
\title{Topology Analysis of the Sloan Digital Sky Survey: \\
   I. Scale and Luminosity Dependence}

\author{Changbom Park\altaffilmark{1}, Yun-Young Choi\altaffilmark{1}, 
Michael S. Vogeley\altaffilmark{2},
J. Richard Gott III\altaffilmark{3}, Juhan Kim\altaffilmark{1}, 
Chiaki Hikage\altaffilmark{4}, 
Takahiko Matsubara\altaffilmark{4}, 
Myeong-Gu Park\altaffilmark{5},
Yasushi Suto\altaffilmark{6}, \& 
David H. Weinberg\altaffilmark{7} }
\author{(For the SDSS collaboration)}

\begin{abstract}

We measure the topology of volume-limited galaxy samples selected from
a parent sample of 314,050 galaxies in the Sloan Digital Sky Survey
(SDSS), which is now complete enough to describe the fully
three-dimensional topology and its dependence on galaxy properties.
We compare the observed genus statistic $G(\nu_f)$ to predictions for
a Gaussian random field and to the genus measured for mock surveys
constructed from new large-volume simulations of the $\Lambda$CDM
cosmology. In this analysis we carefully examine the dependence of the
observed genus statistic on the Gaussian smoothing scale $R_G$ from 3.5 to 11
$\hMpc$ and on the luminosity of galaxies over the range
$-22.50<M_r<-18.5$. The void multiplicity $A_V$ is less than unity at
all smoothing scales. Because $A_V$ cannot become less than 1 through
gravitational evolution, this result provides strong evidence for
biased galaxy formation in low density environments.  We also find
clear evidence of luminosity bias of topology within the
volume-limited sub-samples.  The shift parameter $\Delta \nu$
indicates that the genus of brighter galaxies shows a negative shift
toward a ``meatball'' (i.e. cluster-dominated) topology, while faint
galaxies show a positive shift toward a ``bubble''
(i.e. void-dominated) topology. The transition from negative to
positive shift occurs approximately at the characteristic absolute
magnitude $M_{r*}=-20.4$.  Even in this analysis of the largest galaxy
sample to date, we detect the influence of individual large-scale
structures, as the shift parameter $\Delta \nu$ and cluster
multiplicity $A_C$ reflect (at $\sim 3\sigma$) the presence of the
Sloan Great Wall and a {\large $x$}-shaped structure which runs for
several hundred Mpc across the survey volume.

\end{abstract}
%]

\altaffiltext{1}{Korea Institute for Advanced Study, Dongdaemun-gu, Seoul 130-722, Korea}
\altaffiltext{2}{Department of Physics, Drexel University, 3141 Chestnut Street, Philadelphia, PA 19104, USA}
\altaffiltext{3}{Department of Astrophysical Sciences, Peyton Hall, Princeton University, Princeton, NJ 08544-1001, USA}
\altaffiltext{4}{Department of Physics and Astrophysics, Nagoya University, Chikusa, 
Nagoya 464-8603, Japan}
\altaffiltext{5}{Department of Astronomy and Atmospheric Sciences, Kyungpook National University, Daegu 702-701, Korea}
\altaffiltext{6}{Department of Physics, School of Science, University of Tokyo, Tokyo 113-0033, Japan}
\altaffiltext{7}{Department of Astronomy, 140 W. 18th Ave., Ohio State
University, Columbus, OH 43210, USA}


\section{Introduction}

Topology analysis was introduced in cosmology as a method to test
Gaussianity of the primordial density field as predicted by many
inflationary scenarios (Gott, Melott, \& Dickinson 1986).  The
statistics of the initial density field are thought to be well
preserved at large scales where structures are still in the linear
regime. Therefore, to achieve the original purpose of topology
analysis one needs to use large observational samples and explore the
galaxy density field at large smoothing scales. This requires an
accurate map of the large-scale distribution of galaxies over scales
of several hundred megaparsecs, as is now available from the Sloan
Digital Sky Survey (SDSS).

On smaller scales in the non-linear regime, 
the topology of the galaxy distribution
yields strong constraints on the galaxy formation mechanisms and the
background cosmogony.  Galaxies of various species are distributed in
different ways in space, and the differences can be quantitatively
measured by topology analysis.  By studying galaxy biasing as revealed
in statistics beyond the two-point correlation function and power
spectrum, the complex nature of galaxy formation can be better
understood. The topology statistics can be precision measures of the
galaxy formation process (Park, Kim, \& Gott 2005).  To examine topology at
small scales it is necessary to use dense galaxy redshift samples
which are also large enough in volume to not to be significantly
affected by sample variance.  Using the SDSS, it is now possible to study the
topology of the galaxy distribution down to a Gaussian smoothing scale
of $3 \hMpc$ scale in volume-limited samples that include galaxies as
faint as absolute magnitude $M_{r} = -18$, while still maintaining
reasonably large sample volumes.

Previous topological analysis of SDSS galaxy distribution include the
2D genus (Hoyle {\etal} 2002), 3D genus with Early Data Release
(Hikage {\etal} 2002), and Minkowski Functionals with Sample 12
(Hikage {\etal} 2003). The present paper updates the previous results
with the latest SDSS sample, describe below as Sample 14.
For the first time, we are able to detect the
quantitative signature of luminosity-dependent biasing by
characterizing the genus curves in terms of the statistical
quantities, $\Delta\nu$, $A_V$, and  $A_C$.
$\Delta \nu$ is shift parameter  and
$A_V$ and $A_C$ are cluster and void abundance parameters, respectively.

In this paper we adopt the genus statistic as a measure of the
topology of the smoothed galaxy number density field.  To study the
impact of galaxy biasing, we limit our attention to small-scale
topology, over a range of Gaussian smoothing scales of from 3.5 to 11
$\hMpc$. We examine the scale-dependence of topology to see if there
are differences with respect to the $\Lambda{\rm CDM}$ model.  We also
detect the luminosity bias in the topology of large scale structure.
In section 2 we briefly describe the SDSS and define our
volume-limited SDSS samples. In section 3 we define the topology
statistics and describe our genus analysis procedure.  Section 4
describes our new N-body simulation, which we use for constructing
mock surveys and testing for systematic effects.  In section 5 we
present results of tests for scale and luminosity dependences of the
observed genus curves. We discuss our findings in section 6.

\section{Observational Data Set}
\subsection{Sloan Digital Sky Survey}
The SDSS (York {\etal} 2000; Stoughton {\etal} 2002; Abazajian {\etal} 2003;
Abazajian {\etal} 2004) is a survey to explore the large scale 
distribution of galaxies and quasars, and their physical properties
 by using a dedicated $2.5 {\rm m}$
telescope at Apache Point Observatory. The photometric
survey, which is expected to be completed in June 2005,
will image roughly $\pi$ steradians of the Northern Galactic Cap 
in five photometric 
bandpasses denoted by $u$, $g$, $r$, $i$, and $z$ centered at 
$3551, 4686,6165, 7481,$ and $8931 \AA $, respectively
by an imaging camera with 54-CCDs (Fukugita {\etal} 1996; Gunn {\etal} 1998).
The limiting magnitudes of photometry at a signal-to-noise ratio of $5:1$
are $22.0, 22.2, 22.2, 21.3$, and
$20.5$ in the five bandpasses, respectively. 
The width of the PSF is $1.4\arcsec$, and the photometric
uncertainties are $2 \%$ rms (Abazajian {\etal} 2004).
Roughly $5\times10^7$ galaxies will be cataloged.

After image processing (Lupton {\etal} 2001; Stoughton {\etal}
2002; Pier {\etal} 2003) and calibration (Hogg {\etal} 2001;
Smith {\etal} 2002),
targets are selected for spectroscopic follow-up
observation. The spectroscopic survey is planned to continue through 2008
as the Legacy survey, and produce about $10^6$ galaxy spectra.
The spectra are obtained by two dual fiber-fed CCD spectrographs.
The spectral resolution is $\lambda/\Delta \lambda\sim 1800$, 
and the {\rms} uncertainty in redshift is $\sim 30$ km/s. 
Mainly due to the minimum distance of $55\arcsec$
between fibers, incompleteness of spectroscopic survey reaches 
about 6\% (Blanton {\etal} 2003a) 
in such a way that regions with high surface densities
of galaxies become less prominent. 
This angular variation of sampling density is accounted for in our analysis.
%There are also three stripes in the Southern Galactic Cap observed
%by SDSS. We will not use the data in those stripes
%because they are not useful for the three dimensional topology analysis.

% I eliminated the abbreviations because they are never used again after this
% paragraph
The SDSS spectroscopy yields three major samples: the main galaxy
sample (Strauss {\etal} 2002), the luminous red galaxy sample
(Eisenstein {\etal} 2001), and the quasar sample (Richards {\etal}
2002).  The main galaxy sample is a magnitude-limited sample with
apparent Petrosian $r$-magnitude cut of
$m_{r,\mathrm{lim}}\approx17.77$ which is the limiting magnitude for
spectroscopy (Strauss {\etal} 2002).  It has a further cut in
Petrosian half-light surface brightness
$\mu_{\rm{R50},\mathrm{limit}}=24.5$ mag/arcsec$^2$.  
% irrelevant for this paper
%The galaxies in
%the LRGs are selected by color-magnitude cuts in $g$, $r$, and $i$,
%and have spectroscopic magnitude limit of $m_r\sim 19.5$.  
More details about the survey can be found at
\url{http://www.sdss.org/dr3/}.

In our topology analysis, we use a large-scale structure sample of the SDSS
from the NYU Value-Added Catalog (VAGC, Blanton {\etal} 2004).
%The galaxies in VAGC are within $2\arcsec$ of an MGS, LRGS, or quasar target, 
%within $2\arcsec$ of an SDSS spectrum, or 
%pass the MGS-like criteria (more inclusive than MGS selection).
As of the writing of this paper, the most up-to-date large-scale
structure sample is Sample 14, which covers 3,836 square degrees
of the sky, and contains 314,050 galaxies between redshift of $0.001$
and $0.5$, surveyed as of November 2003.  The large-scale structure
sample also comes with an angular selection function of the survey
defined in terms of spherical polygons (Hamilton \& Tegmark 2004),
which takes into account the incompleteness due to mechanical
spectrograph constraints, bad spectra, or bright foreground stars.
 
\subsection{Sample Definitions for Genus Analysis}
To study the three-dimensional topology of the smoothed galaxy number density
distribution, it is advantageous for the observational sample
to have the lowest possible surface-to-volume ratio.
For this reason we trim Sample 14 as shown
in Figure~\ref{fig1}, where the solid lines delineate our sample
boundaries in the survey coordinate plane ($\lambda,\eta$).
We discard the three southern stripes and the small areas 
protruding or isolated from the main surveyed regions. 
%\begin{figure}
%%\epsscale{1.2}
%\plotone{f1.eps}
%\caption{Angular definition of the SDSS sample used for our
%topology analysis. Solid lines delineate the boundaries of the analysis
%regions in the survey coordinate plane $(\lambda, \eta)$.
%}
%\label{fig1}
%\end{figure}
These cuts decrease the number of galaxies
from 314,050 to 239,216 in two analysis regions.
Within our sample boundaries, 
we account for angular variation of the survey completeness
by using the angular selection function provided with the large scale
structure sample data set.
To facilitate our analysis, we make two arrays of square pixels
of size $0.025\degr \times 0.025\degr$
in the ($\lambda,\eta$) sky coordinates which cover the two
analysis regions, and store the angular selection function calculated
by using the {\texttt mangle} routine (Hamilton \& Tegmark 2004). 
At the location of each pixel, 
the routine calculates the survey completeness in a spherical polygon
formed by the adaptive tiling algorithm (Blanton {\etal} 2003a)
used for the SDSS spectroscopy. The resulting useful area within the analysis 
regions with non-zero selection function is 0.89 steradians.

Analysis region 1 contains the famous ``Sloan Great Wall'' for which
redshift slices are shown by Gott {\etal} 2003.  Figure~\ref{fig2}
shows galaxies with $14.5 \leq m_r \leq 17.5$ in two $7.5\degr$-thick
slices in the analysis region 2.  We assume a flat $\Lambda{\rm CDM}$
cosmology with $\Omega_{\rm m} = 0.3$ to convert redshifts to comoving
coordinates.  In the upper slice of Figure~\ref{fig2} there is a weak
wall of galaxies that extends over $\sim 700\hMpc$ at comoving
distance of roughly $r=400\hMpc$.  Void, wall, and filamentary
structures of galaxies are seen through these slices.  A roughly
spherical void of size $\sim 100\hMpc$ in diameter is seen in the
slices at distance of about $200\hMpc$.  In the lower slice there is
a $\sim 300\hMpc$ size structure which looks like a runner or a
{\large $x$} mark, formed by several neighboring voids of various
sizes as was the structure in the CfA slice (Geller \& Huchra 1989).
We note several small voids nested within larger ones.

%\begin{figure*}
%\epsscale{0.80}
%\plotone{f2a.eps}
%\plotone{f2b.eps}
%\caption{Distribution of galaxies with $14.5 \leq m_r \leq 17.5$ 
%in two contiguous $7.5\degr$-thick slices in the analysis region 2. 
%The radial coordinate is
%Comoving distance and 
%angular coordinate is SDSS survey longitude $\lambda$.
%[{\it See the electronic edition of the Journal for a color version
%of this figure. 
%Galaxies brighter and fainter than $M_* = -20.44$ are distinguished by color.}]
%}
%\label{fig2} \end{figure*}

In our topology analysis we use only volume-limited samples of
galaxies defined by absolute magnitude limits.  Figure~\ref{fig3}
shows galaxies in Sample 14 in redshift-absolute magnitude space.  The
smooth curves delineate the sample boundaries corresponding to our
choice of apparent magnitude limits of $14.5 \leq m_r \leq 17.5$,
after correction for Galactic reddening (Schlegel, Finkbeiner, \&
Davis {\etal} 1998).  The faint limit of $m_r=17.5$ is slightly
brighter than the spectroscopic selection criterion of $m_r<17.77$, to
allow use of some early data that used a brighter limit. The
bright-end apparent magnitude limit of $m_r=14.5$ is imposed to avoid
small incompleteness that is caused by the exclusion of galaxies with
large central surface brightness (to avoid spillover in the
spectrograph CCDs) and associated with the quality
of deblending of large galaxies.
%\begin{figure*}
%%\epsscale{1.1}
%%\plotone{f3.eps}
%\caption{Sample definitions in the redshift-absolute magnitude space.
%Top panel shows boundaries of three volume-limited samples used for studying
%scale dependence of topology. In the bottom panel four samples used for 
%luminosity bias study are defined.
%The smooth curves delineate the sample boundaries corresponding to
%our choice of apparent magnitude limits of $14.5 \leq m_r \leq 17.5$.
%[{\it See the electronic edition of the Journal for a color version
%of this figure.}]
%}\label{fig3}
%\end{figure*}
The most natural volume-limited sample is the one containing the
maximum number of galaxies.  We vary the faint and bright absolute
magnitude limits to find such a sample and label this our ``Best''
sample.  It is defined by a absolute-magnitude limits $-21.53 \leq
M_r < -20.15$, which correspond to a comoving distance range of
$162.9<r< 319.0 \hMpc$ or redshift range $0.055 <z<0.109$ when the
apparent magnitude cut is applied.  The comoving distance and
redshift limits are obtained by using the formula
\begin{equation}
m_r - M_r = 5 {\rm log} (r(1+z)) + 25 + K(z),
\end{equation}
where $K(z)$ is the K-correction and $r(1+z)$ is the luminosity
distance. 
We use a polynomial fit to the mean K-correction within $0 <z <0.3$,
\begin{equation}
K(z)=2.3537(z-0.1){^2}+1.04423(z-0.1)-2.5 {\rm log} (1+0.1).
\end{equation}
The rest-frame absolute magnitudes of galaxies in Sample 14 are computed
in fixed bandpasses, shifted to $z=0.1$, using Galactic reddening corrections
and K-corrections (for a full
description, see Hogg {\etal} 2002 and Blanton {\etal} 2003b).
This means that galaxies at $z=0.1$ have K-correction of
$-2.5 {\rm log}(1+0.1)$, independent of their SEDs.
We do not take into account galaxy evolution effects.
The definition of the `Best' sample is shown in Figure~\ref{fig3}
and Table~1.

The upper panel of
Figure~\ref{fig3} also shows the absolute-magnitude and redshift
limits for two additional samples, which we label ``Sparse'' and
``Dense'' because of their mean density relative to the ``Best''
sample. The Dense sample has a bright absolute magnitude limit just
below the faint limit of the Best sample. The faint limit of the Dense
sample is determined by maximizing the number of galaxies.  For the
Sparse sample, the bright absolute magnitude limit was chosen to be
fainter than $-22.22$, above which galaxies in Sample 14 are missing at
far distances. The faint limit of the Sparse is determined by
maximizing the number of galaxies contained in the sample. In what follows
these samples are used to study dependence of topology on scale.

The lower panel of Figure~\ref{fig3} shows the absolute-magnitude and
redshift limits of four samples that we use to study the luminosity
dependence of topology.  Each sample has an absolute magnitude range
of two magnitudes. The brightest sample, L1, has a bright magnitude limit
of $M_r=-22.5$.  Each luminosity sample is further divided into three
subsamples defined as in Table~1.  Each subsamples has half the number
of galaxies contained in its parent sample, and the brightest
subsample does not overlap with the faintest one in absolute
magnitude. Because the subsamples occupy the same volume of the
universe and contain the same number of galaxies, differences among
them are free from sample-variance and Poisson fluctuation; any
variation in topology is purely due to difference in the absolute magnitude 
of the galaxies.

\section{Theory for Topology Analysis}
\subsection{Genus and its Related Statistics}
The genus is a measure of the topology of isodensity
contour surfaces in a smoothed galaxy density field. It is defined as
\begin{equation}
G = {\rm Number~of~holes~in~contour~surfaces-
Number~of~isolated~regions}
\end{equation}
in the isodensity surfaces at a given threshold level.
The Gauss-Bonnet theorem connects the global topology with an integral
of local curvature of the surface $S$, {\ie}
\begin{equation}
G={1 \over {4 \pi}} \int_{S} \kappa da,
\end{equation}
where $\kappa$ is the local Gaussian curvature.
In the case of a Gaussian field the genus per volume as a
function of density threshold level is known (Doroshkevich 1970; Adler 1981; Hamilton, Gott, \& Weinberg 1986):
\begin{equation}
g(\nu)=g(0) (1-\nu^{2}) e^{-\nu^{2}/2},
\end{equation}
where $\nu \equiv (\rho-{\bar\rho})/\sigma$ is the threshold density in unit of
standard deviations $\sigma=\langle(\rho-{\bar\rho})^2\rangle^{1/2}$
from the mean. The amplitude is 
$$ g(0) = {1 \over {(2 \pi)^2}} ({{\langle k^2\rangle} \over 3})^{3/2}, $$
where 
$${\langle k^2\rangle} = \int P(k) k^2 d^3 k / \int P(k) d^3 k $$ depends on
the power spectrum $P(k)$ of the smoothed density field.  To separate
variation in topology from change of the the one-point density
distribution we measure the genus as a function of the volume-fraction
threshold $\nu_f$. This parameter defines the density contour surface
such that the volume fraction in the high density region is the same
as the volume fraction in a Gaussian random field contour surface with
a value of $\nu = \nu_f$.  Any deviation of the observed genus curve
from the Gaussian one is evidence for non-Gaussianity of the primordial
density field and/or that acquired due to the nonlinear gravitational
evolution or galaxy biasing.  There have been a number of studies on
the effects of non-Gaussianity on the genus curve (Weinberg {\etal}
1987; Park \& Gott 1991; Park, Kim, \& Gott 2005).

To measure the deviation of the genus curve from Gaussian, several
statistics have been suggested.  First is the amplitude drop,
$R=G_{\rm obs}/G_{\rm G}$ where $G_{\rm obs}$ is the amplitude of the
observed genus curve and $G_{\rm G}$ is that of a Gaussian field which
has the observed power spectrum (Vogeley {\etal} 1994).  It is a
measure of the phase correlation produced by the initial
non-Gaussianity, the gravitational evolution, and the galaxy biasing.
We will not calculate $R$ in this paper and defer it to later works in
which we will examine the genus over a larger range of smoothing
scales that include the linear regime.  Here we simply measure the
observed genus amplitude $G_{\rm obs}$ to test for scale and
luminosity dependence. $G_{\rm obs}$ is measured by finding the best
fitting Gaussian genus curve over $-1\leq\nu\leq 1$.

The shift parameter $\Delta\nu$ is defined as
\begin{equation}
\Delta\nu = \int d\nu G_{\rm obs}(\nu)\nu/\int d\nu G_{\rm fit}(\nu),
\end{equation}
where the integral is over $-1\leq\nu\leq 1$, and $G_{\rm obs}$
and $G_{\rm fit}$ are the observed and the best-fit Gaussian genus curves
(Park {\etal} 1992).
It measures the threshold level where topology is maximally sponge-like.
For a density field dominated by voids, $\Delta\nu$ is positive and we say
that the density field has a ``bubble-like'' topology.
For a cluster dominated field, $\Delta\nu$ is negative and we say that 
the field has a ``meatball-like'' topology.

The abundances of clusters and voids relative to that expected for a Gaussian
random field are measured by the parameters $A_{\rm C}$ and $A_{\rm V}$.
These parameters are defined by
\begin{equation}
A=\int d\nu G_{\rm obs}(\nu)/\int d\nu G_{\rm fit}(\nu), 
\end{equation}
where the integration intervals are $+1.2 <\nu_f< +2.2$ for $A_{\rm
C}$ and $-2.2 <\nu_f< -1.2$ for $A_{\rm V}$ (Park, Gott, \& Choi 2001;
Park, Kim, \& Gott 2005). These intervals are centered near the minima
of the Gaussian genus curve ($\nu=\pm\sqrt{3}$) and stay away from the
thresholds where the genus curve is often affected by the shift
phenomenon.  These ranges also exclude extreme thresholds where for
low density regions the volume-fraction threshold level $\nu_f$ is
very sensitive to the density value.  These parameters are defined so
that $A_{\rm C, V}>1$ means that more independent clusters or voids are
observed than predicted by a Gaussian field at fixed volume fraction,
whereas $A_{\rm C, V}<1$ means that fewer independent clusters or voids
are seen.  A detailed study of the effects of the gravitational
evolution, the galaxy biasing, and the cosmogony on the $\Delta\nu$,
$A_{\rm V}$, and $A_{\rm C}$ statistics is presented by Park, Kim, \&
Gott (2005).

\subsection{Analysis Procedure}

\begin{figure*}
%\epsscale{1.2}
\vspace{5cm}
\plotone{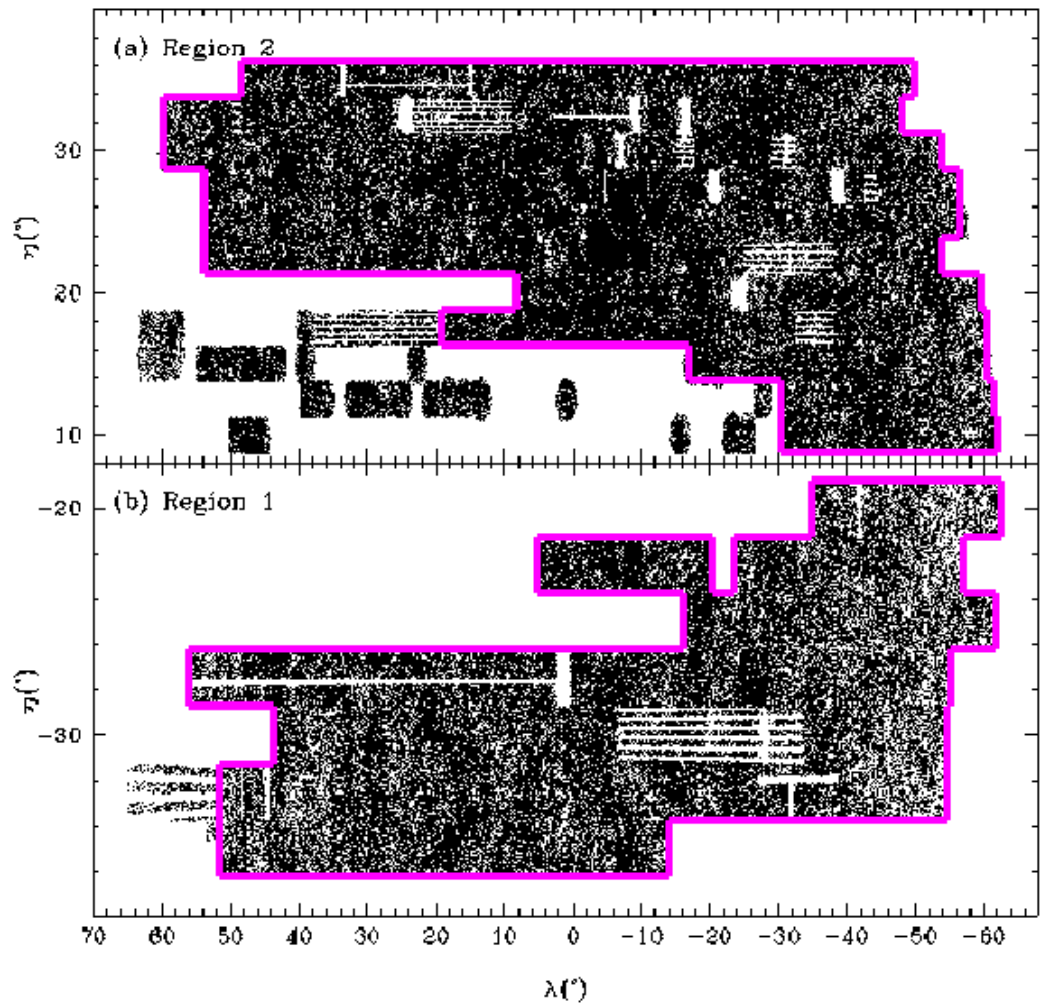}
\caption{Angular definition of the SDSS sample used for our
topology analysis. Solid lines delineate the boundaries of the analysis
regions in the survey coordinate plane $(\lambda, \eta)$.
}
\label{fig1}
\end{figure*}

\begin{figure*}
\vspace{4.5cm}
\epsscale{0.8}
\plotone{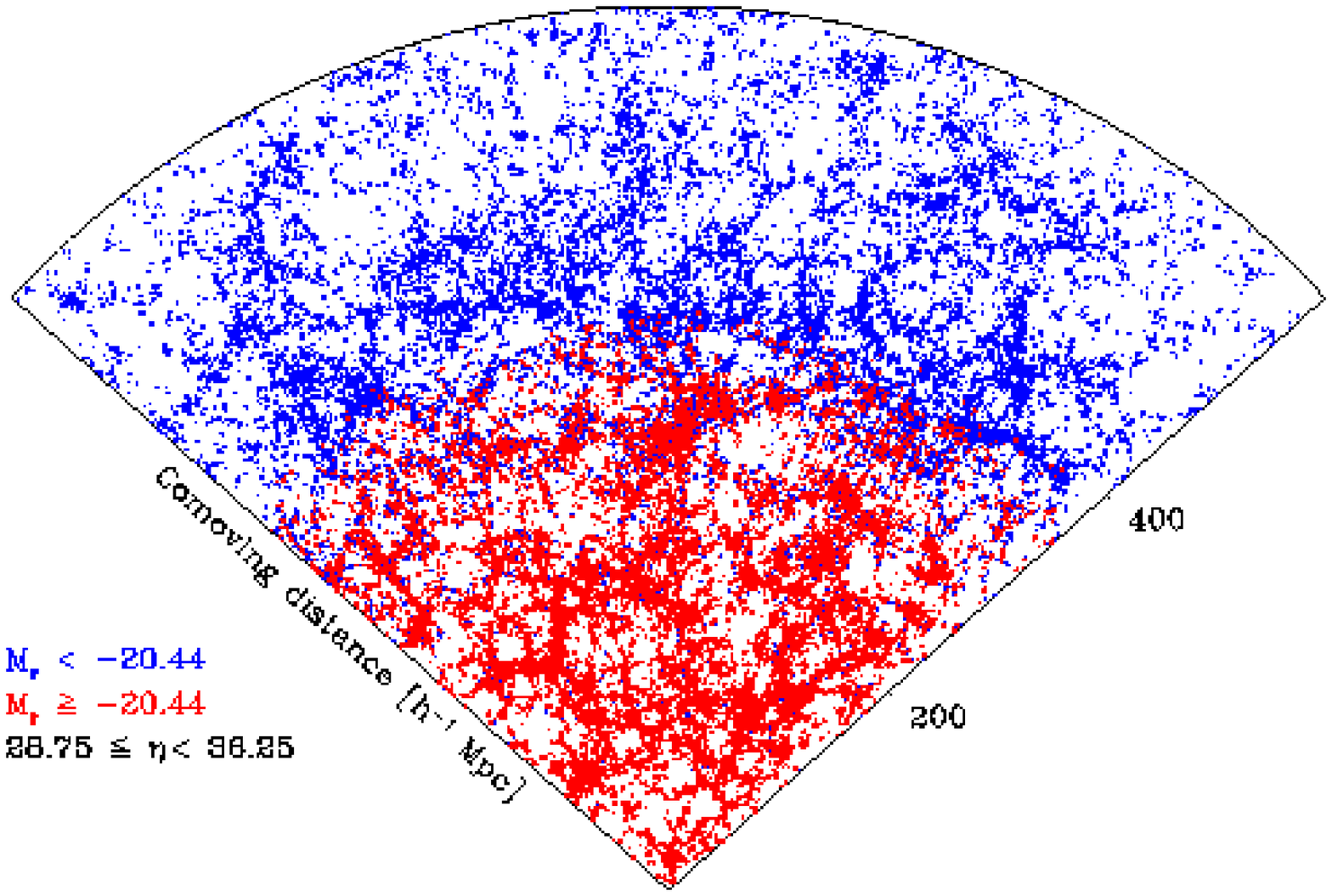}
\vspace{1.5cm}
\plotone{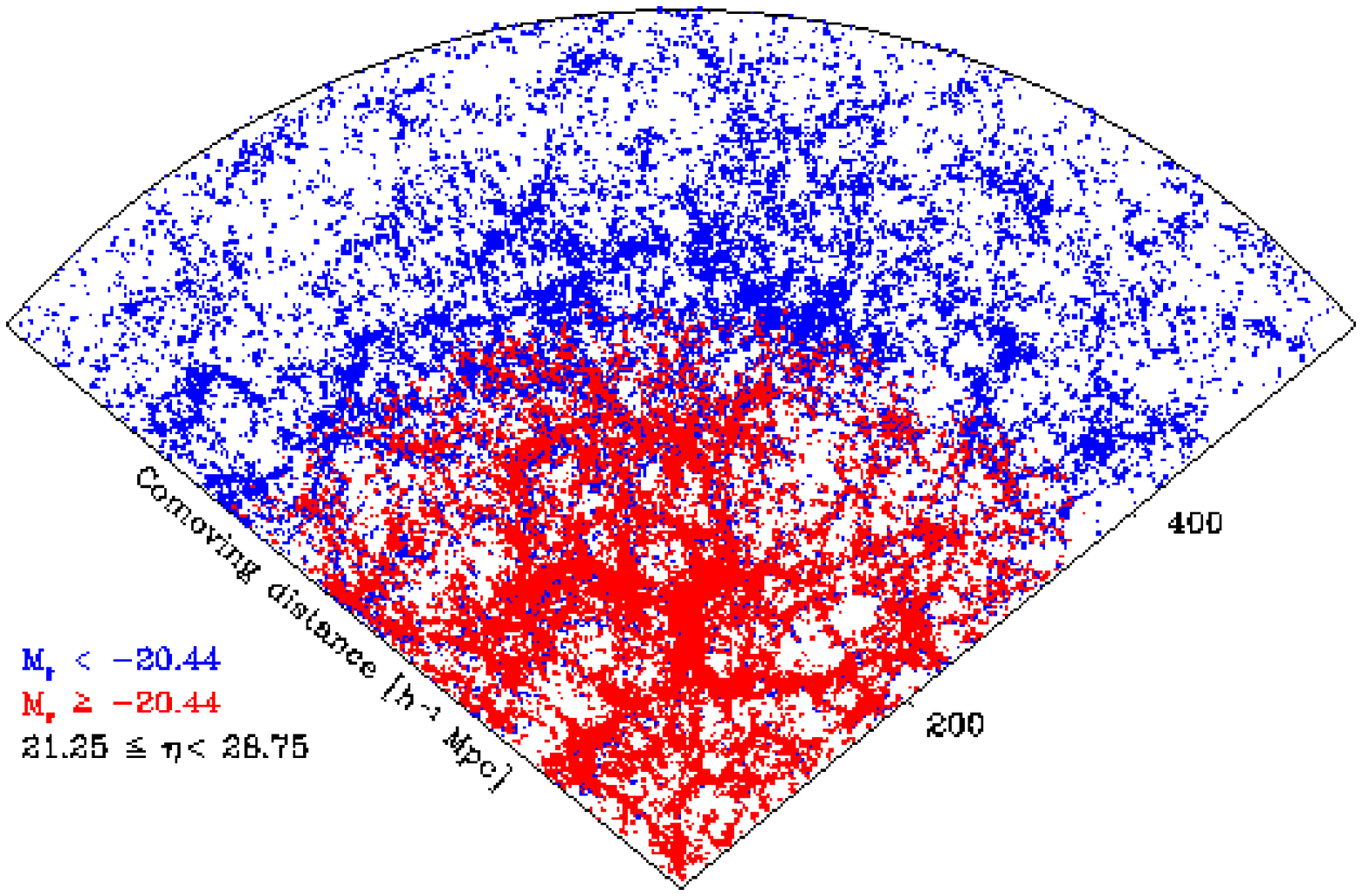}
\caption{Distribution of galaxies with $14.5 \leq m_r \leq 17.5$
in two contiguous $7.5\degr$-thick slices in the analysis region 2.
The radial coordinate is
Comoving distance and
angular coordinate is SDSS survey longitude $\lambda$. 
[{\it See the electronic edition of the Journal for a color version
of this figure. 
Galaxies brighter and fainter than $M_* = -20.44$ are distinguished by color.}]
}
\label{fig2} \end{figure*}

\begin{figure*}
%\epsscale{1.1}
\vspace{4.5cm}
\plotone{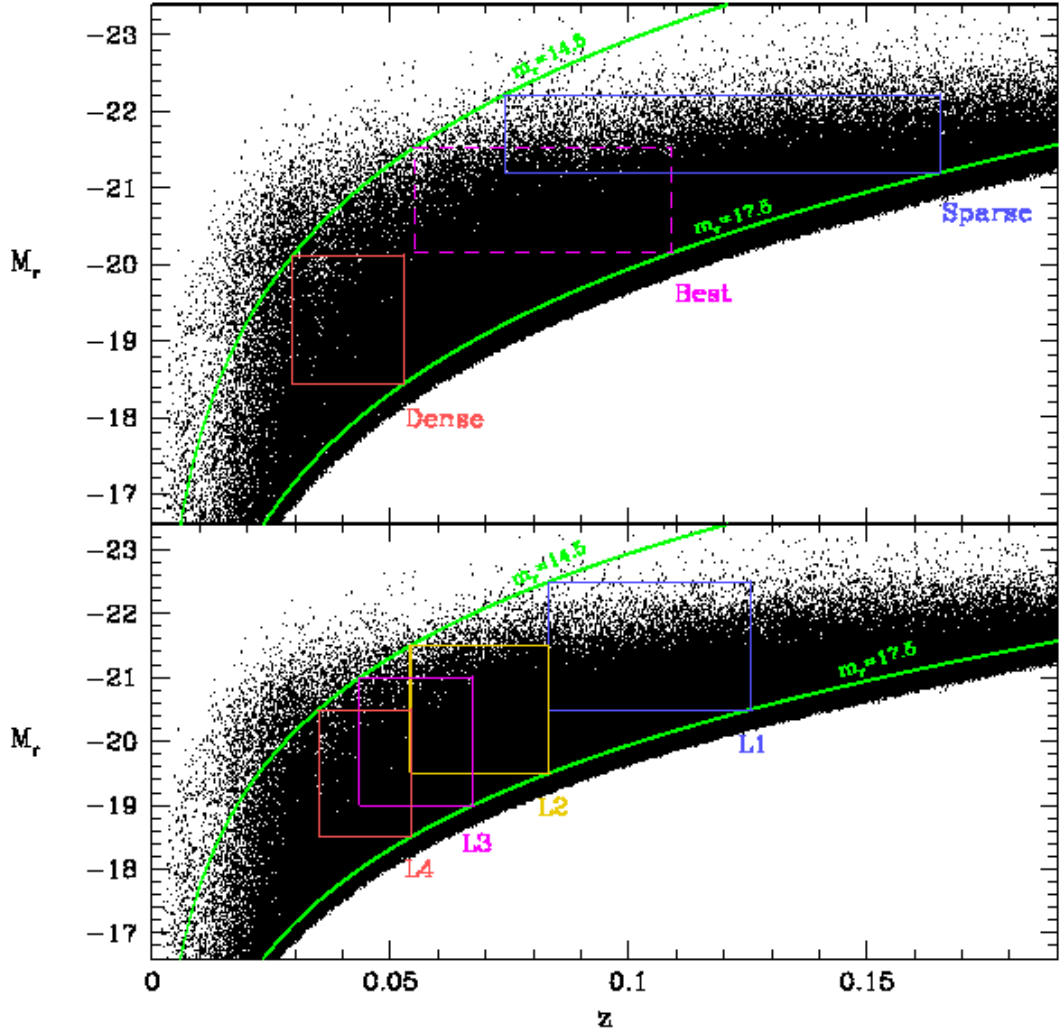}
\caption{Sample definitions in the redshift-absolute magnitude space.
Top panel shows boundaries of three volume-limited samples used for studying
scale dependence of topology. In the bottom panel four samples used for
luminosity bias study are defined.
The smooth curves delineate the sample boundaries corresponding to
our choice of apparent magnitude limits of $14.5 \leq m_r \leq 17.5$.
[{\it See the electronic edition of the Journal for a color version
of this figure.}]
}\label{fig3}
\end{figure*}

\begin{figure*}
%\begin{minipage}{180mm}
\vspace{3.cm}
\begin{center}
\includegraphics[scale=0.5]{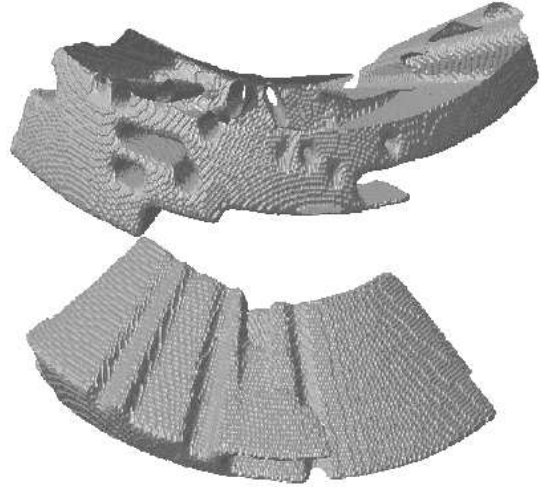}
\end{center}
\caption{Three-dimensional views of the mask array looking toward the Earth
from the far side after smoothing and trimming.
This mask is used for analysis of the Best sample.
The upper piece is the region 2, and the lower one is the region 1.}\label{fig4}
%\end{minipage}
\end{figure*}

\begin{figure*}
\vspace{-2cm}
%\epsscale{0.6}
%\plottwo{fig5a_res200.ps}{fig5b_res200.ps}\\
%\plottwo{fig5c_res200.ps}{fig5d_res200.ps}
\begin{center}
\includegraphics[scale=0.25]{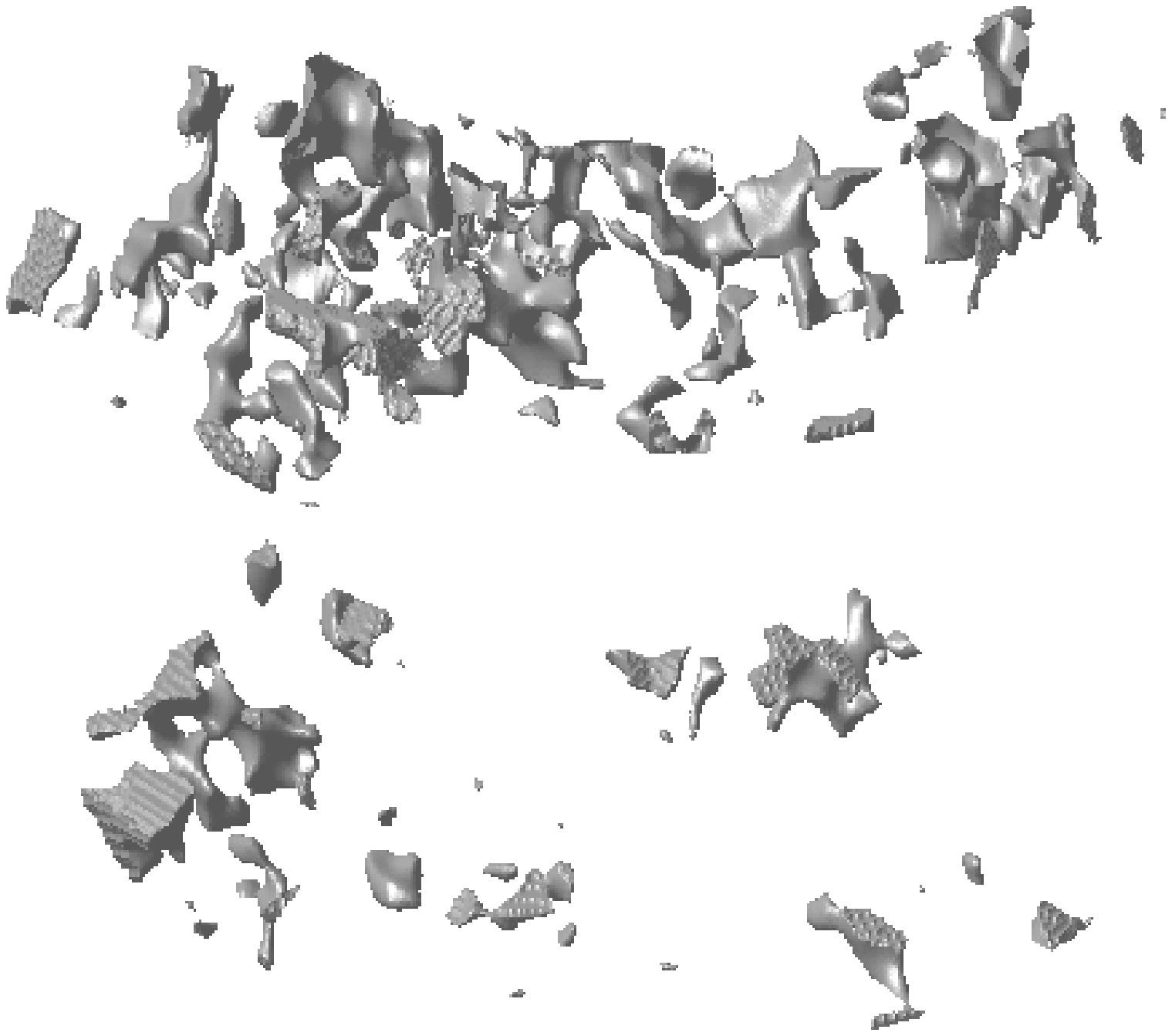}
\includegraphics[scale=0.25]{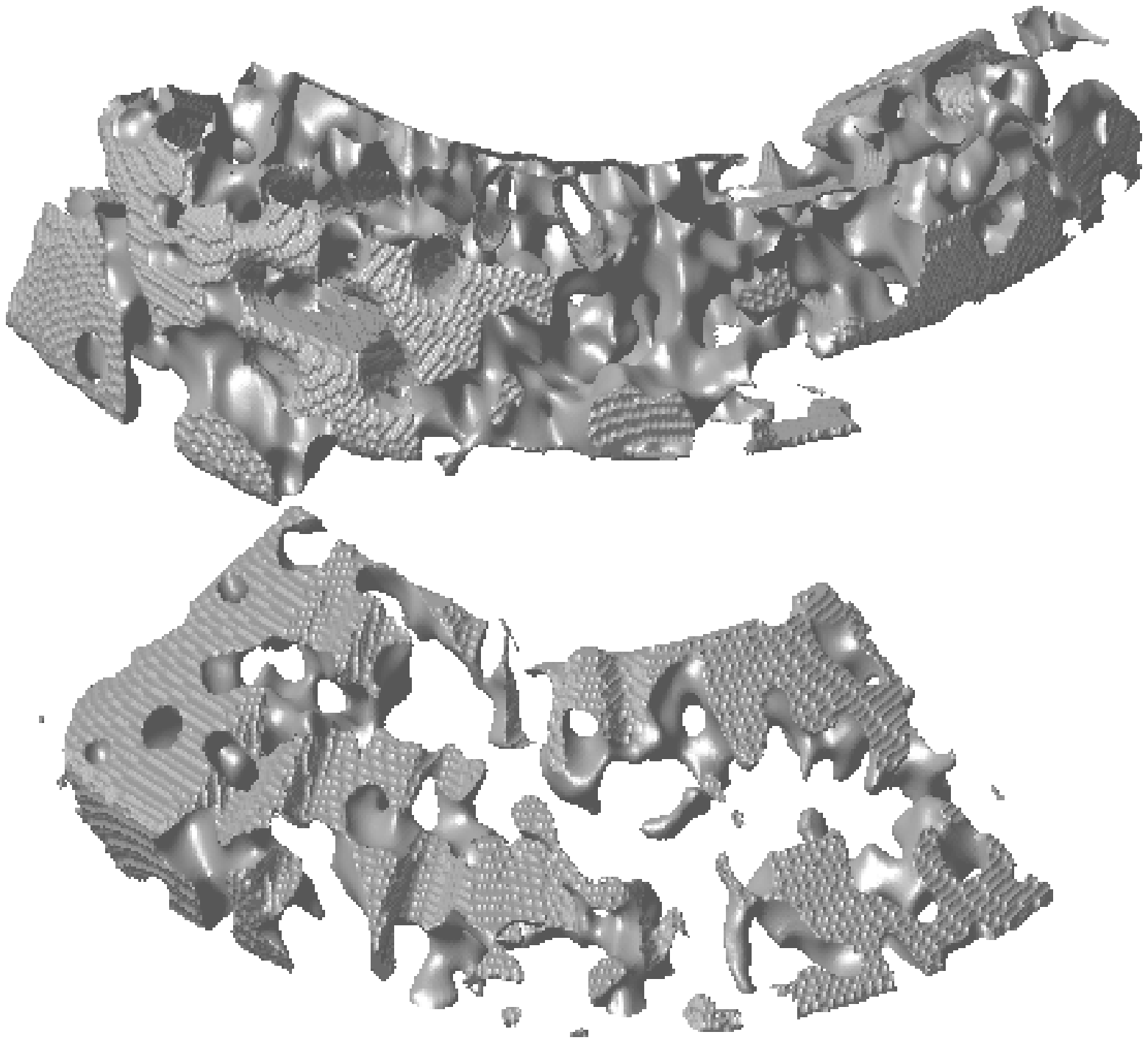}
\includegraphics[scale=0.25]{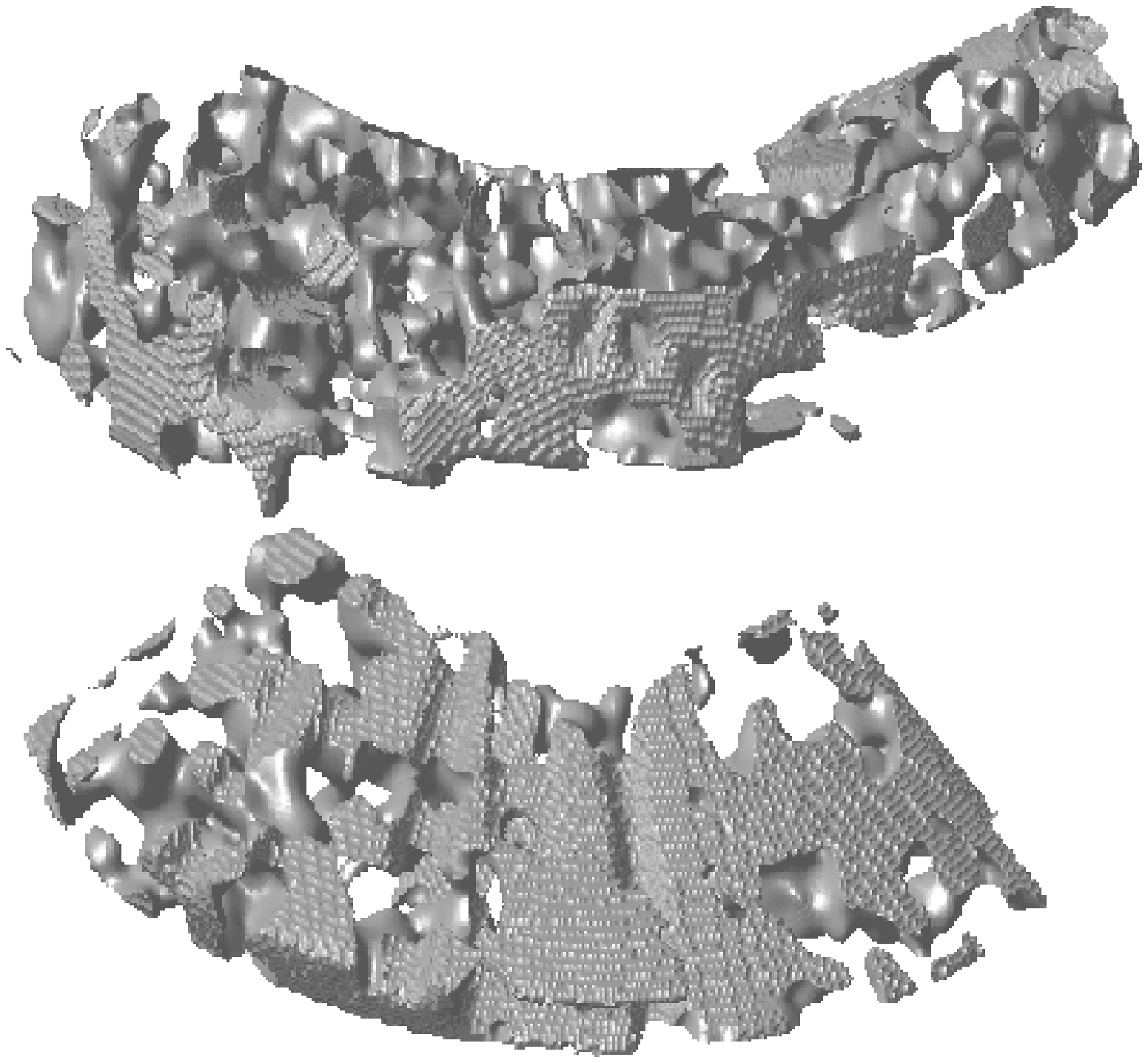}
\includegraphics[scale=0.25]{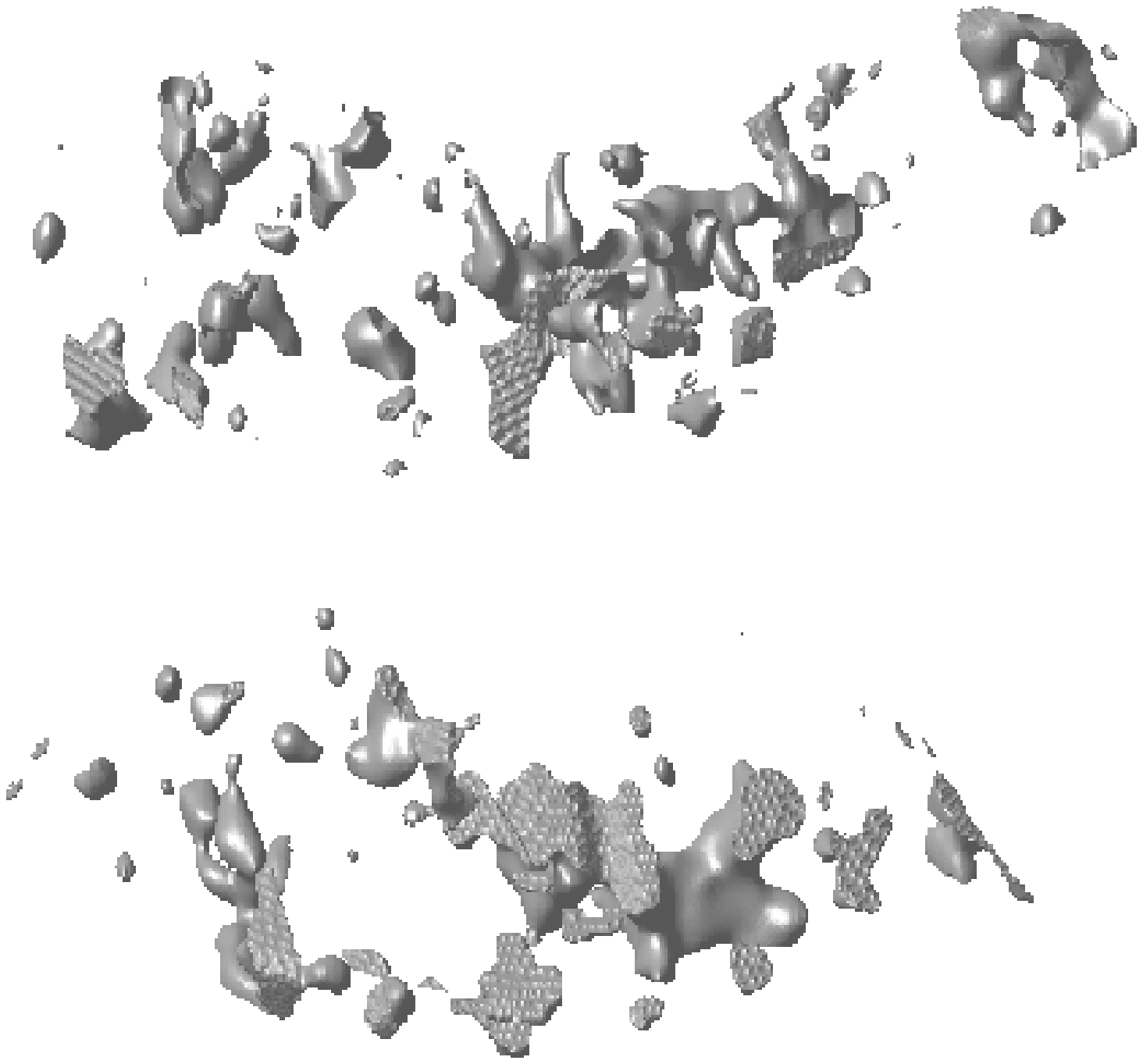}
\end{center}
\caption{Three-dimensional view of the galaxy number density field of
the Best sample after smoothing and trimming. On the left, density contours
enclose low density regions occupying $7\%$ and $50\%$ of the volume, and
on the right, contours enclose high density regions filling $50\%$ and $7\%$
of the volume of the Best sample.
}\label{fig5}
%\end{minipage}
\end{figure*}

To measure the genus using the {\tt contour3D} algorithm (Weinberg
1988), we prepare an estimate of the smoothed density field on a grid,
with pixels that lie outside the survey region
flagged. Figure~\ref{fig1} together with the angular selection
function table defines our angular mask.  Table~1 lists the distance
range for each sample.  Using the angular mask and distance limits, we
make a three-dimensional mask array of size $512^3$ for each sample.
The pixel size is always restricted to be slightly smaller than
$R_{\rm G}/3$ where $R_{\rm G}$ is the Gaussian smoothing length. The
Earth is located at the center of one face of the cubic array that
forms the $x$-$z$ plane. This mask array contains zeros in pixels that
lie outside the boundaries shown in Figure~\ref{fig1}. Mask array
pixels that lie within the boundaries are assigned a selection
function value read off from the angular selection function
table. Because that table consists of fine pixels of $1.5\arcmin$
size, the mask array faithfully represents the survey selection
effects. We also make a $512^3$ galaxy density array into which we bin
the SDSS galaxies which fall inside our sample boundaries.  After a
Gaussian smoothing length $R_{\rm G}$ is chosen, both the mask array
and the galaxy density array are smoothed and divided by each other in
the sense of $\rho_{\rm g}/\rho_{\rm mask}$.  This yields an array of
smoothed galaxy density estimates that accounts for both the angular
selection function and the effect of the survey boundary. In regions
of the ratio array for which the corresponding value in the smoothed
mask array value is smaller than $0.69$, we flag the ratio array with
negative values to indicate that they are outside the analysis
region. This removes not only regions that are formally outside the
survey volume, but also regions that are too close to the survey
boundary, where the signal-to-noise ratio of the estimated density
field is low. We choose a threshold value of $0.69$ as a compromise
between homogeneous smoothing and a larger survey volume.  This value
corresponds to the value of a smoothed mask at a distance of $0.5
R_{\rm G}$ inside an infinite planar survey boundary.  
Figure~\ref{fig4} shows the mask array for the Best sample, 
after smoothing and trimming.
%\begin{figure*}
%%\begin{minipage}{180mm}
%\begin{center}
%\includegraphics[scale=0.5]{f4.eps}
%%\includegraphics[scale=0.45]{fig4b_res200.ps}
%\end{center}
%\caption{Three-dimensional views of the mask array looking toward the Earth
%from the far side after smoothing and trimming.
%This mask is used for analysis of the Best sample.
%The upper piece is the region 2, and the lower one is the region 1.}\label{fig4}
%%\end{minipage}
%\end{figure*}
%Figure~\ref{fig5} shows the Best sample smoothed in the way explained above.
%\begin{figure*}
%%\epsscale{0.6}
%%\plottwo{fig5a_res200.ps}{fig5b_res200.ps}\\
%%\plottwo{fig5c_res200.ps}{fig5d_res200.ps}
%\begin{center}
%\includegraphics[scale=0.25]{f5d.eps}
%\includegraphics[scale=0.25]{f5c.eps}
%\includegraphics[scale=0.25]{f5a.eps}
%\includegraphics[scale=0.25]{f5b.eps}
%\end{center}
%\caption{Three-dimensional view of the galaxy number density field of
%the Best sample after smoothing and trimming. On the left, density contours
%enclose low density regions occupying $7\%$ and $50\%$ of the volume, and
%on the right, contours enclose high density regions filling $50\%$ and $7\%$
%of the volume of the Best sample.
%}\label{fig5}
%%\end{minipage}
%\end{figure*}
The two figures on the left of Figure~\ref{fig5} 
show under-dense regions at volume fractions
of $7\%$ and $50\%$. Those on the right show overdense regions. 
The Sloan Great Wall is visible in Region 1 at $7\%$ high density level.

After the smoothed galaxy density field is obtained, we compute the
genus at each volume fraction $\nu_f$ using {\texttt contour3d}. We
then estimate the best fit Gaussian amplitude $G_{\rm obs}$ and the
other genus-related parameters, $\Delta\nu$, $A_{\rm V}$, and $A_{\rm
C}$ for each genus curve $G(\nu_f)$.

\section{Mock Surveys from Simulations}

To measure the topology statistics accurately and to detect any weak
non-Gaussianity or dependence of topology on the physical properties
of galaxies, accurate modeling of the survey and the analysis are
required to eliminate systematic effects.  For this purpose we have
made a new large N-body simulation of a $\Lambda{\rm CDM}$ universe,
which has the mean particle number density much higher than that of
the galaxies in the SDSS 
and at the same time can safely contain the large-scale modes that
modulate the density field over the maximum scales explored by SDSS.
We adopt the cosmological parameters measured by the WMAP (Spergel
{\etal} 2003), which are $\Omega_{\rm m}=0.27, \Omega_{\rm b}=0.0463,
\Omega_{\rm \Lambda}=0.73, h=0.71$ and $\sigma_8 = 0.9$. 
Here, $\sigma_8$ is the {\rms} 
fluctuation of mass in a 8 $\hMpc$ radius spherical tophat. 
The physical size of the simulation cube is 5632 $\hMpc$
which is much larger than any volume-limited SDSS sample we use here.
The simulation follows the evolution of 8 billion 
$= 2048^3$ CDM particles whose initial
conditions are laid down on a $2048^3$ mesh.
We have used a new parallel PM+Tree N-body code (Dubinski {\etal} 2004)
to increase the spatial dynamic range.
The gravitational force softening parameter is set to $0.1$ times the mean
particle separation. The
particle mass is $1.6\times 10^{12} h^{-1} M_{\sun}$ and the mean
separation of particles is 2.75 $\hMpc$
while that of SDSS galaxies in our Best sample is about $6 \hMpc$.
The simulation was started at $z=17$ and followed the gravitational evolution
of the CDM particles at $170$ time steps using $128$ CPUs on the IBM p690+
supercomputer of the Korea Institute of Science and Technology Information.
%The mass of a particle in the higher resolution simulation is 
%$9.4 \times 10^9 h^{-1} M_{\sun}$ and dark halos with mass higher
%than $5\times 10^{11} h^{-1} M_{\sun}$ can be almost completely identified.
%The mean separation of those halos turns out to be about $5 \hMpc$
%while that of SDSS galaxies in our Best sample is about $6 \hMpc$.
%The simulation was started at $z=17$ and followed the gravitational evolution
%of the CDM particles at $680$ time steps using $128$ CPUs on a IBM p690+
%supercomputer of Korea Institute of Science and Technology Information.

We make 100 mock surveys in the 5632 $\hMpc$ simulation for each of
our samples and subsamples and for each smoothing scale in both real
and redshift spaces. The total number of mock surveys is 3600.  We use
these mock surveys to estimate the uncertainties and systematic biases
in the measured genus and its related statistics.  We randomly locate
`observers' in the 5632 $\hMpc$ simulation at $z=0$ and make
volume-limited surveys as defined in Table~1.  The number of galaxies
in each mock survey is constrained to be almost equal to that of each
observational sample.  We analyze the resulting mock samples in
exactly the same way that the observational data are analyzed.

Any systematic bias due to the finite number of galaxies or smoothing
effects should also appear in the results of analysis of mock samples.
Variation of the genus among the mocks provides an estimate of the
random uncertainties of the observations.  We compare the mean genus
and genus-related statistics over 100 mock samples in real space with
those from the whole simulation cube.  Differences between the mocks
and full cube indicate systematic biases for which we then correct the
observed values.  With the exception of the plotted genus curves, we
correct all results in this fashion.  Note that we use the mock
surveys in real space to estimate the systematic biases because it is
only in the real space where the true values of the topology
parameters can be measured by using the whole simulation data. Sets of
100 mock surveys made in redshift space are used to estimate the
uncertainties in the observed genus and its related statistics.  The
cosmological parameters used in our simulation are only slightly
different from those applied to observational data. Hence, assuming
the WMAP cosmological parameters are approximately correct, systematic
bias correction factors and uncertainty limits should be correctly
estimated from mock surveys.

\section{Genus Results}

\subsection{Overview}

For each of the volume-limited samples of SDSS listed in Table~1, we
compute the genus at $501$ volume-fraction threshold levels spaced by
$\delta \nu_f=0.01$.  The shortest smoothing length applied is set to
about $(2 {\rm ln}2)^{-1/2}\approx 0.85$ times the average intergalaxy
separation ${\bar d}=\bar{n}^{-1/3}$, which corresponds to a Gaussian
smoothing kernel whose FWHM is $2{\bar d}$.

To estimate the uncertainties of these measurements, in each case we
use the variance among 100 mock surveys drawn from the the $5632\hMpc$
$\Lambda{\rm CDM}$ simulation in redshift space.  These uncertainties
include the effects of both Poisson fluctuations and sample variance.
Figures~\ref{fig6} and \ref{fig8} show the genus curves and uncertainties.
The smooth curve in each plots is the mean over 100 mock samples.
To see the dependence of our results on the observer's location 
we have made additional mock surveys at locations where the local 
overdensity smoothed over $8\hMpc$ tophat is between 0 and 1, 
peculiar velocity is $600\pm50$ km/sec, and the peculiar velocity shear 
is $|v - {\bar v}| / {\bar v} < 0.5$. 
Here $\bar v$ is the bulk velocity of the $8\hMpc$ sphere 
around the particle (cf. Gorski {\etal} 1989). 
The resulting genus statistics are of little difference 
with those from random locations.

We then measure the amplitude $G_{\rm obs}$ of the best-fit Gaussian
genus curve, the shift $\Delta \nu$, and the cluster and void
multiplicity parameters $A_{\rm C}$ and $A_{\rm V}$ for each
curve. Using results from the mock surveys, we correct these
parameters for systematic effects that result from the shape of the
survey volume (see Section 4).  Figures~\ref{fig7} and \ref{fig9}
present these parameters and compare them with results from mock
surveys of the $\Lambda$CDM simulation. Table~2 lists all the measured
parameters, both corrected (in parentheses) and uncorrected for
systematics.  Note that the figures plot the genus per
smoothing volume $gR_G^3=(G_{\rm obs}/V_{\rm survey})R_G^3$
rather than $G_{\rm obs}$.  It can be seen in Table~2 that the systematic biases
are smaller for cases with smoothing lengths shorter than the mean galaxy
separation of a given sample.

In the following sections, we examine the dependence of the genus
on both smoothing scale and luminosity of galaxies. Note that we
test for luminosity bias using subsamples that cover the same
physical volume (same angular and comoving distance limits), so that
there are no sample variance effects.

\subsection{Scale Dependence}

To test for dependence of the genus parameters on smoothing scale we
begin by examining the Best sample (see Table~1 and
Figure~\ref{fig3}).  Genus curves measured from the Best sample at
smoothing scales $R_G=5, 6$ and $8\hMpc$ are shown in
Figure~\ref{fig6}a.  The genus-related statistics with systematic biases
corrected are shown in
Figure~\ref{fig7} (the middle 5 points) and summarized in Table~2, for
five smoothing lengths, $R_G=5, 6, 7, 8$ and $9\hMpc$.

The top panel of Figure~\ref{fig7} shows the genus density per 
smoothing volume $gR_{G}^3$. The middle panel shows the shift
parameter $\Delta \nu$. The lower panel shows the cluster and void
multiplicity parameters $A_{\rm C}$ (filled symbols) and $A_{\rm V}$
(open symbols), respectively.  The shaded areas indicate $1\sigma$
limits calculated from 100 mock surveys from the $\Lambda{\rm CDM}$
simulation.  In the lower panel, shaded areas are not shown for $A_V$.

% amplitude of Best

In the top panel of Figure~\ref{fig7} we find that the genus per
smoothing volume $gR_G^3$ slightly rises with smoothing scale. This
trend is as expected. For a simple power law spectrum of density
fluctuations, the genus density per smoothing volume is
proportional to $(n+3)^{1.5}$ (Melott, Weinberg, \& Gott 1988), where
$n$ is the power index of power spectrum, $P(k) \propto k^{n}$, of
galaxy distribution. However, because the CDM power spectrum has a maximum at a
larger wavelength than any smoothing scale that we apply, we expect to
measure higher genus density per smoothing volume as we increase $R_G$
in the case of the $\Lambda{\rm CDM}$ model.

% shift of Best

The middle panel of Figure~\ref{fig7} shows that the shift parameter
for the Best sample is negative, $\Delta\nu<0$, and is well below the
$\Lambda$CDM prediction. On a smoothing scale of $6\hMpc$, the
probability of observing a lower value of $\Delta\nu$ in the
$\Lambda$CDM model is $P=0.02$. Thus, the Best sample exhibits a
strong meatball (cluster-dominated) shift.

% cluster, void multiplicity of Best

The cluster multiplicity $A_{\rm C}$ for the Best sample is
consistently below unity (see lower panel of Figure~\ref{fig7}) and
below the $\Lambda$CDM prediction, indicating that there are fewer
independent isolated high density regions than for a Gaussian random
field or the $\Lambda$CDM model. The probability of finding a lower
value of $A_{\rm C}$ with a smoothing scale of $6\hMpc$ in the
$\Lambda$CDM model is only $P=0.03$.

The strong meatball shift ($\Delta\nu<0$) and low cluster multiplicity
($A_{\rm C} <1$) in the Best sample are probably caused by the Sloan
Great Wall in Region 1 and the {\large $x$}-shaped structure in Region
2.  The wall is located at the distance between about $160$ and $240
\hMpc$ and is almost fully contained in the Best sample.

We find that the void multiplicity parameter, $A_V$, is much lower
than 1 at all smoothing scales explored (see the bottom panel of
Figure~\ref{fig7}).  This implies that voids are more connected than
expected for Gaussian fields.  Park, Kim, \& Gott (2005) find that
non-linear gravitational evolution causes the $A_V$ parameter to rise,
but that $A_V$ can become lower than unity when a proper prescription
for biased galaxy formation is applied.  Thus, the observation that
$A_{V} <1$ is strong evidence for biased galaxy formation. This
measurement of small-scale topology provides a new quantitative test
for galaxy formation theories. Here we find that the scale dependence of
genus indicates that $A_V$ only weakly depends on the smoothing scale.
In contrast with the observed void multiplicity,
the $A_V$ parameter of the $\Lambda{\rm CDM}$ matter field (not shown
in Figure~\ref{fig7}) is greater than
1 at scales smaller than about $9 \hMpc$ (Park, Kim, \& Gott 2005).  
Although the mock surveys are constructed by treating all matter
particles as candidate galaxies and their topology is not to be
directly compared with that of observed galaxies, it still provides some 
guide line.
Note that in the early topology analysis dark matter particles are
often used for this comparison (Canavezes {\etal} 1998; Protogeros \&
Weinberg 1997). Simple prescriptions like
the peak biasing scheme are also often used (Park \& Gott 1991; 
Vogeley {\etal} 1994; Colley {\etal} 2000).
When the purpose of topology analysis is to discriminate among different
galaxy formation mechanisms using observational data, one should apply
proper prescriptions for identifying galaxies in the N-body simulation.

%In order to address the statistical significance of the Sloan Great Wall
%more realistically, we use the dark halos identified from our higher resolution%simulation. Dark halos are found at locations of local density peaks, and 
%matter particles are iteratively assigned to a dark halo 
%located at a density peak when they are 
%gravitationally bound. When particles are bound to more than one dark halos,
%they are accepted as members of subhalos only when they are within the tidal
%radius. Our halo finding algorithm is described elsewhere (Kim \& Park 2005). 
%Even though the algorithm is superior to the conventional Friend-of-Friend
%algorithm, the center of massive clusters is occupied by big dark halos and
%there are numerous small dark halos. We use the Halo Occupation Distribution 
%(HOD) modeling to populate `galaxies' within dark halos. 
%The HOD parameters given in Table~3 of Zehavi {\etal} (2004) are used 
%to obtain the minimum halo mass above the absolute magnitude threshold and 
%the mean number of satellite substructures above the threshold.
%Even though the cosmological parameters are slightly different, the number
%density (mean separation $\bar d=6.5 \hMpc$) of `galaxies' assigned
%in accordance with the HOD prescription is close to that ($\bar d=6.1 \hMpc$)
%of the Best sample.
%We have made 100 mock surveys corresponding to the Best sample. 

% Results for Sparse, Dense

For comparison to the Best sample, we also examine the genus curves of
the Sparse and Dense samples, as shown in Figure~\ref{fig6}b at two
smoothing scales, plotted together with the mean genus curves from 100
mock surveys. The parameters $gR_G^3$, $\Delta\nu$, $A_{\rm C}$, and
$A_{\rm V}$ for the Dense sample are the leftmost three
points in each panel of Figure~\ref{fig7}, while
the rightmost two points in each panel show results for the Sparse
sample.

% amplitude

The genus amplitude of the Sparse sample is significantly lower than
that of the $\Lambda{\rm CDM}$ model.  Rather than being a real signal
we think this result has been caused by lack of bright galaxies at far
distances in Sample 14.  The number density of galaxies in
volume-limited samples defined by absolute magnitude limits, starts to
radially decrease when the absolute magnitude limit exceeds
$-22$. Paucity of bright galaxies can be easily noticed in Figure 3 at
absolute magnitudes $M_r \leq -22.5$ and redshift $z\geq 0.13$ (see
also Figure 3 of Tegmark {\etal} 2004 who took into account the
evolution of galaxies).  The lower density of galaxies at the far side
of the Sparse sample 
effectively decreases the sample volume at a given threshold level
and, thus decreases the amplitude of the genus curve.

% shift

The Dense sample shows a clear positive shift $\Delta \nu >0$ when
compared with the matter distribution of the $\Lambda{\rm CDM}$ model.
%({\bf \tt add more description....}).
This shift is caused mainly by the luminosity bias that will be
described below in Section~5.3. 

% cluster, void multiplicity?

In both the Dense and Sparse samples we find that the cluster
multiplicity $A_{\rm C}$ is somewhat larger than for the $\Lambda$CDM
simulation. But $A_{\rm C}$ of the Sparse sample seems 
overestimated due to the radial density drop, which causes individual
galaxies at the far side appear as isolated high density regions.
The void multiplicity $A_{\rm V}$ for the Dense and Sparse samples is
less than unity, in agreement with the results for the Best
sample. Therefore, we find a low value of $A_{\rm V}$ in all of these
samples.

Within each of the samples (Best, Dense, Sparse), we find no clear
evidence for scale dependence of the parameters $\Delta\nu$, $A_{\rm
C}$ or $A_{\rm V}$. As discussed above, the increase of $gR_G^3$ with
smoothing scale is as expected for a CDM-like power spectrum.

\subsection{Topology as a Function of Galaxy Luminosity}

It is well-known that the strength of galaxy clustering depends on
luminosity (Park {\etal} 1994; Norberg {\etal} 2001; Zehavi {\etal}
2004; Tegmark {\etal} 2004).  Park {\etal} (1994) demonstrated that
this luminosity bias is due to bright galaxies which tend to populate
only dense regions, and completely avoid voids.  To study the galaxy
luminosity bias beyond one-point and two-point statistics (\eg, Jing
\& B\"{o}rner 2003 and Kayo {\etal} 2004 for three-point correlation
function; Hikage {\etal} 2003 for Minkowski functionals), here we test for
luminosity bias in the topology of the large-scale structure of
galaxies. 

%\begin{figure}
%\epsscale{0.65}
%\plotone{f8a.eps}
%\plotone{f8b.eps}
%\caption{($a$) Genus curve calculated by using all
%galaxies contained in each luminosity sample. 
%Smoothing length is roughly 0.85 times the mean separation between
%galaxies in each luminosity sample.  
%($b$) Genus curves of luminosity subsamples
%which have half the number of galaxies contained in their parent
%luminosity samples. Only one subsample for each luminosity sample is
%shown. Smoothing lengths are 7.5, 5.0, 5.0, and 4.4 $\hMpc$ for L1-1,
%L2-2, L3-3, and L4-3 subsamples, respectively.
%[{\it See the electronic edition of the Journal for a color version
%of this figure.}]
%}\label{fig8}
%\end{figure}

For this analysis we construct four luminosity samples L1, L2, L3, and
L4, each of which spans an absolute magnitude range $\Delta M_r=2$
(see Figure~\ref{fig3} and Table~1 for definitions). Figure~\ref{fig8}
presents the genus curves calculated for these luminosity
samples. Table~2 lists the measured genus-related statistics. The
smoothing lengths are set to be about 0.85 times the mean separation
between galaxies.  The genus curves averaged over 100 mock luminosity
samples are also plotted as smooth curves. We estimate uncertainties
in the genus curves using the variance among the mock samples.

% amplitude - nothing new to say? I moved the theoretical reference to above

% shift

% cluster, void multiplicity

The genus curve for the brightest sample, L1, shows a clear negative
shift $\Delta \nu<0$ with respect to the mock survey result, while the
faintest sample, L4, shows a clear positive shift $\Delta \nu>0$.
After systematic biases are corrected, using the mock surveys in real
space, we find the negative shift for the sample L1 is reduced
(see values in parentheses in Table~2) and the fainter samples show
even stronger positive shifts. The large significance of the change
from negative (or zero) shift in bright samples to a positive genus
shift in fainter samples seems to indicate luminosity bias effects on
topology.  However, we must be cautious, because these samples cover
different physics volumes; it might be the case that this trend is
produced by local variation in topology.

To remove this ambiguity we divide each luminosity sample into three
subsamples that cover the same volume of space but with different
absolute magnitude limits and with exactly half the number of galaxies
in the parent sample (see Table~1 for definitions).  For example, the
brighter upper half of galaxies in the luminosity sample L1 is called
L1-1, and the fainter half L1-3.  The middle subsample L1-2 overlaps
with the brighter and fainter ones in absolute magnitude.  Again, we
set the smoothing lengths to about 0.85 times the galaxy mean
separation.  The measured statistics are listed in Table~2 and plotted
in Figure~\ref{fig9}.  In Figure~\ref{fig9} each subsample appears
as one point at the median absolute magnitude of galaxies in the
subsample.

%\begin{figure}
%\epsscale{0.9} 
%\plotone{f9.eps} 
%\caption{Genus-related statistics for luminosity subsamples (see Table~1 for
%definitions). The measured values of subsamples that belong to the
%same luminosity sample are connected together except for the $A_V$
%parameter. The smoothing lengths adopted are $R_G=$ 7.5, 5.0, 5.0, and
%4.4 $\hMpc$ for subsamples of L1, L2, L3, and L4, respectively. Shaded
%regions are the 1 $\sigma$ variation regions calculated from
%100 mock surveys.
%In the bottom panel, $A_C$ is given by filled symbols, $A_V$
%by open symbols, and the shaded areas are shown only for
%$A_C$.
%[{\it See the electronic edition of the Journal for a color version
%of this figure.}]
%}
%\label{fig9}
%\end{figure} 

The differences among the subsamples drawn from a luminosity sample
must be purely due to the luminosity bias, apart from variations
caused by Poisson fluctuations. The luminosity bias of topology is
most clearly detected in the case of the shift parameter. In all
luminosity samples the faintest subsample has a positive shift
relative to the brightest one (see the middle panel of
Figure~\ref{fig9}).  The trend across the parent luminosity samples is also
consistent with this phenomenon. This trend is particularly obvious in 
the case of samples L2 and L3, for which the smoothing length is the same.  The
transition from positive to negative shift appears to occur at around
the characteristic absolute magnitude for SDSS galaxies, $M_{r*} = -20.44$.

Only the L2 and L3 samples are analyzed at the same smoothing scales, thus
their genus amplitudes can be directly compared.  We find no
statistically significant trend in the genus density among subsamples
within each luminosity sample nor across the luminosity samples.
However, a weak dependence on luminosity can be seen, in the sense
that brighter galaxies have higher genus density.  

We find no systematic trend for the cluster and void multiplicity
parameters, $A_C$ and $A_V$, except that the $A_V$ parameter is somewhat lower
for the faintest luminosity sample, L4.

To see the luminosity bias more clearly, in Figure~\ref{fig8}b we plot
the genus curves of luminosity subsamples whose absolute magnitude
intervals do not overlap with one another.  
The systematic change in the shift of the
genus curves toward positive $\nu$ is evident as the luminosity of
subsamples decreases (see Figure~\ref{fig9}).  
These curves are plotted
alongside genus curves averaged over 100 mock surveys that simulate
the luminosity subsamples.  
Note that the $A_C$
parameter is exceptionally low in case of subsamples of the L2 sample
where the Sloan Great Wall is fully contained.

Our finding that the genus curves for galaxies fainter than $M_*$ tend
to have positive shifts implies that the density field of this class
of galaxies has a bubble-shifted topology ($\Delta \nu >0$).  In other
words, the distribution of faint galaxies, which is less clustered
than that of bright galaxies (as measured by the amplitude of the
two-point correlation function), has empty regions nearly devoid of
faint (as well as bright) galaxies.  Further, we find that underdense
regions are more connected to one another than expected for a Gaussian
field ($A_{V} <1$), and this shift is particularly evident in the
distribution of faint galaxies. 
In contrast, bright galaxies show a meatball shift ($\Delta \nu
\lesssim 0$): they form isolated clusters and filaments surrounded by
large empty space.

%\begin{figure}
%\epsscale{1.1}
%\plotone{f10.eps}
%\caption{Distribution of galaxies in the luminosity sample L2 in the comoving
%distance versus survey longitude coordinate plane projected to the median
%volume latitude. Galaxies are distinguished by color and point type in 
%accordance with their absolute magnitudes. Crosses are the brightest, 
%circles the middle, and dots the faintest.
%[{\it See the electronic edition of the Journal for a color version
%of this figure.}]
%%}\label{fig10}
%\end{figure}
%
\section{Conclusions}

The SDSS is now complete enough to allow us to study the three
dimensional topology of the galaxy distribution and its dependence on
physical properties of galaxies. We analyze large-scale structure
dataset Sample 14 of the NYU Value-Added Galaxy Catalog derived from
the SDSS. In particular, we study the dependence of topology of
the large scale structure on the smoothing scale and the galaxy
luminosity. Even though the angular mask of Sample 14 is still very
complicated, we are able to measure the genus statistic accurately by
making extensive and careful use of mock surveys generated from a new
large volume N-body simulation.

\begin{figure}
\vspace{2cm}
\begin{center} 
\includegraphics[scale=0.48]{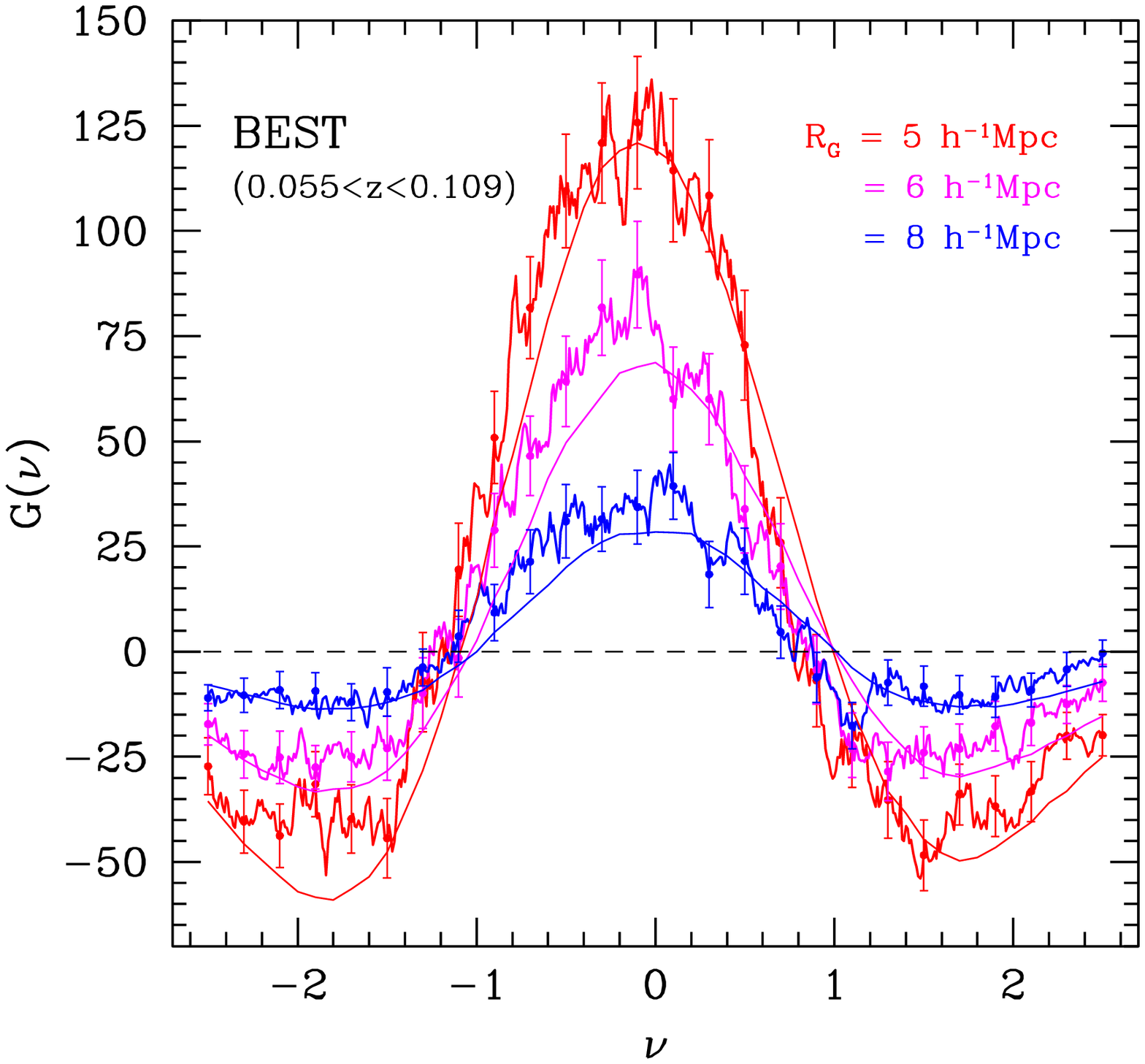}\\
\vspace{-0.5cm}
\includegraphics[scale=0.56]{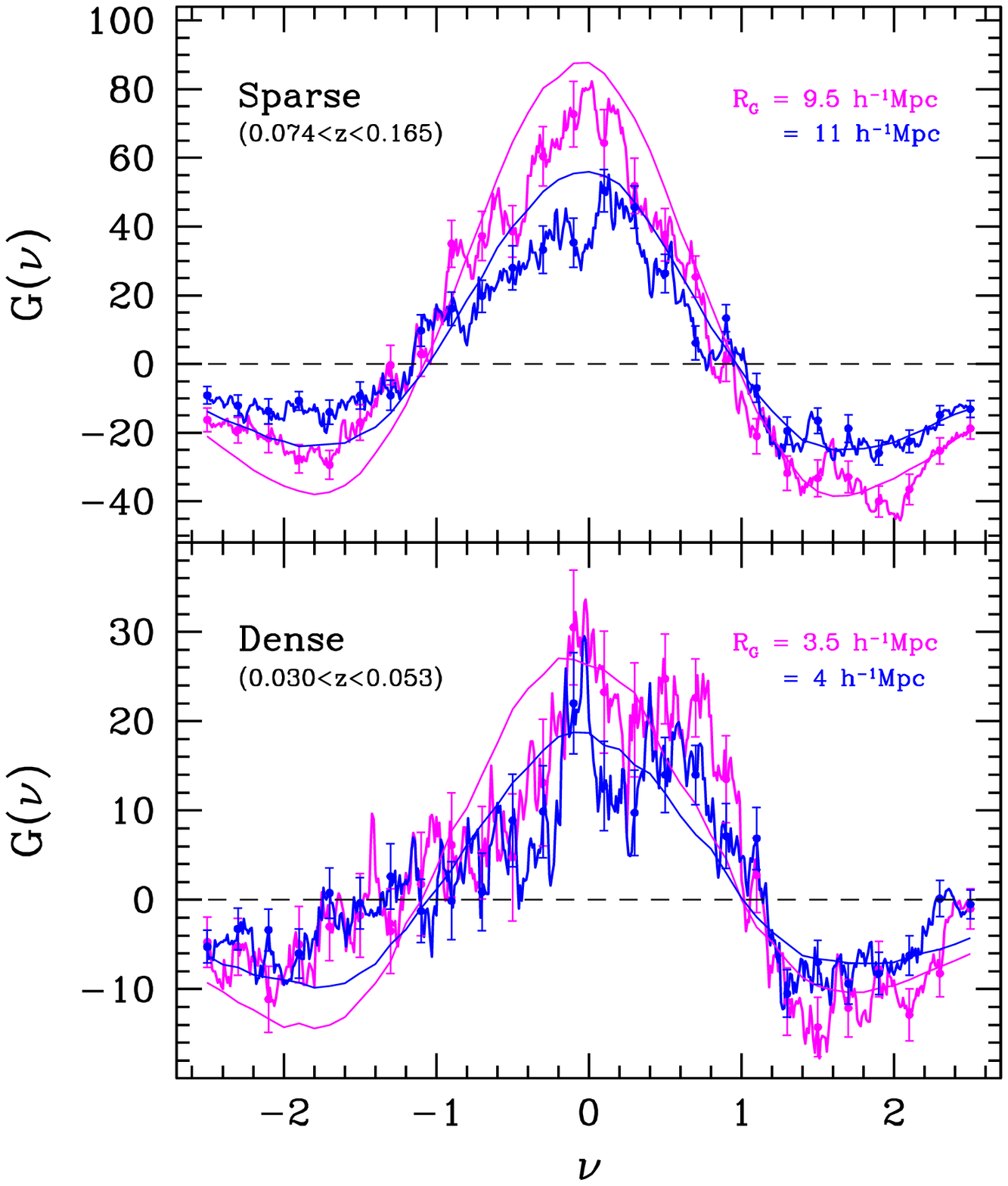}
\caption{($a$) Genus curves measured from the Best sample at three smoothing
scales. Solid curves are the genus curves averaged over 100 mock
surveys in a $\Lambda{\rm CDM}$ simulation in redshift space at the
same smoothing lengths. These curves have not been corrected for
systematic uncertainties. ($b$) Similar genus curves for the Sparse and Dense
samples and corresponding mock surveys.
[{\it See the electronic edition of the Journal for a color version
of this figure.}]
}\label{fig6}
\end{center} 
\end{figure}

Overall, the observed genus curves strongly resemble the random phase
genus curve (see Figure~\ref{fig6}). This supports the idea that the observed
structure arises from random quantum fluctuations as predicted by
inflation.  But on top of this general pattern we observe small
deviations, as might arise from non-linear gravitational evolution
(Melott, Weinberg,
\& Gott 1986; Park \& Gott 1991; Matsubara 1994) and biasing (Gott,
Cen, \& Ostriker 1996).

\begin{figure}
\epsscale{0.90}
\vspace{2cm}
\plotone{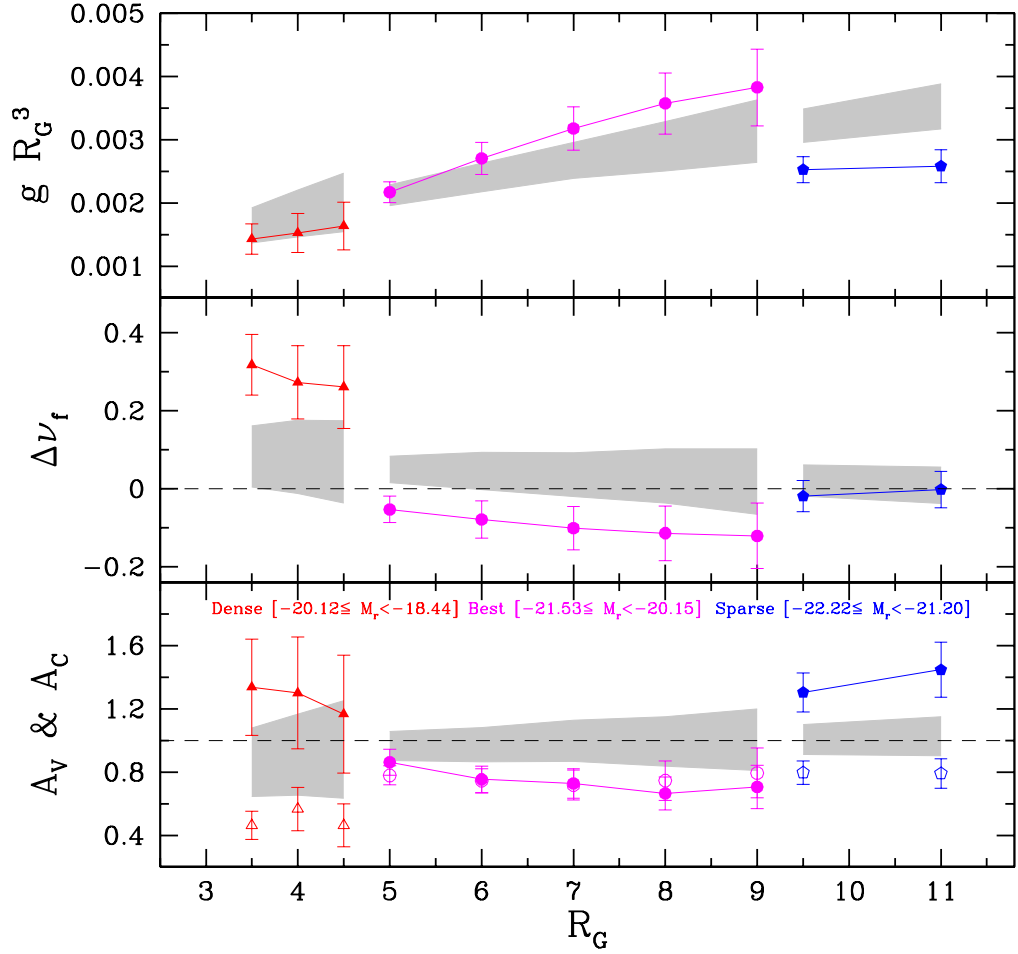}
\caption{Genus-related statistics as a
function of the Gaussian smoothing length, $R_G$. Systematic bias
corrections are made by using mock surveys in real space of the
$\Lambda{\rm CDM}$ model.  Uncertainty limits are derived from mock
surveys in redshift space.  The shaded areas denote the $1 \sigma$
upper and lower limits estimated from the mock samples. In the lower
panel, $A_C$ is given by filled symbols, $A_V$ by open symbols, and
the shaded areas are shown only for $A_C$.
[{\it See the electronic edition of the Journal for a color version
of this figure.}]
}\label{fig7}
\end{figure}

\begin{figure}
\vspace{2cm}
\epsscale{0.65}
\plotone{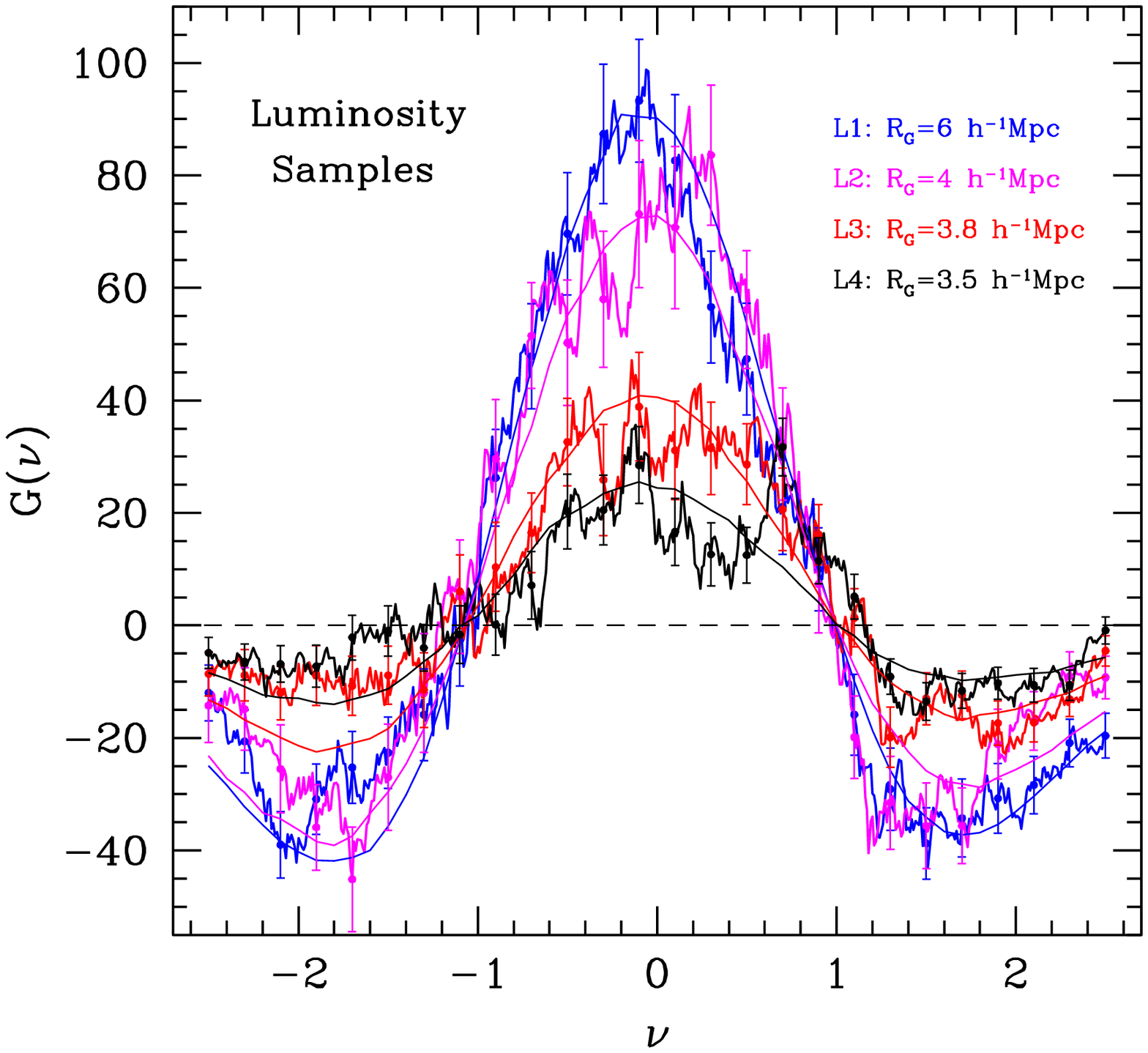}
\plotone{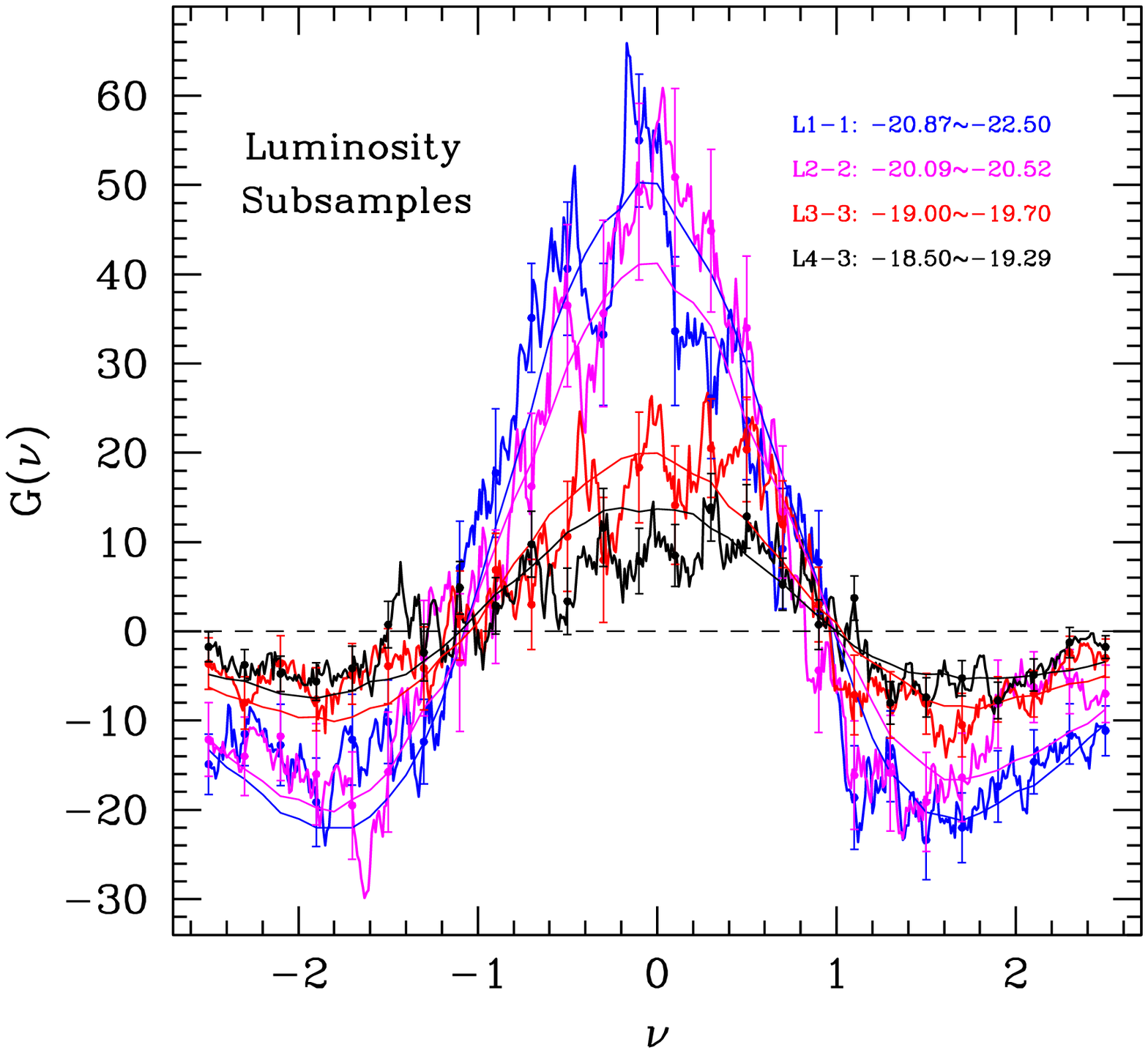}
\caption{($a$) Genus curve calculated by using all
galaxies contained in each luminosity sample.
Smoothing length is roughly 0.85 times the mean separation between
galaxies in each luminosity sample.
($b$) Genus curves of luminosity subsamples
which have half the number of galaxies contained in their parent
luminosity samples. Only one subsample for each luminosity sample is
shown. Smoothing lengths are 7.5, 5.0, 5.0, and 4.4 $\hMpc$ for L1-1,
L2-2, L3-3, and L4-3 subsamples, respectively.
[{\it See the electronic edition of the Journal for a color version
of this figure.}]
}\label{fig8}
\end{figure}

\begin{figure}
\vspace{2cm}
\epsscale{0.9}
\plotone{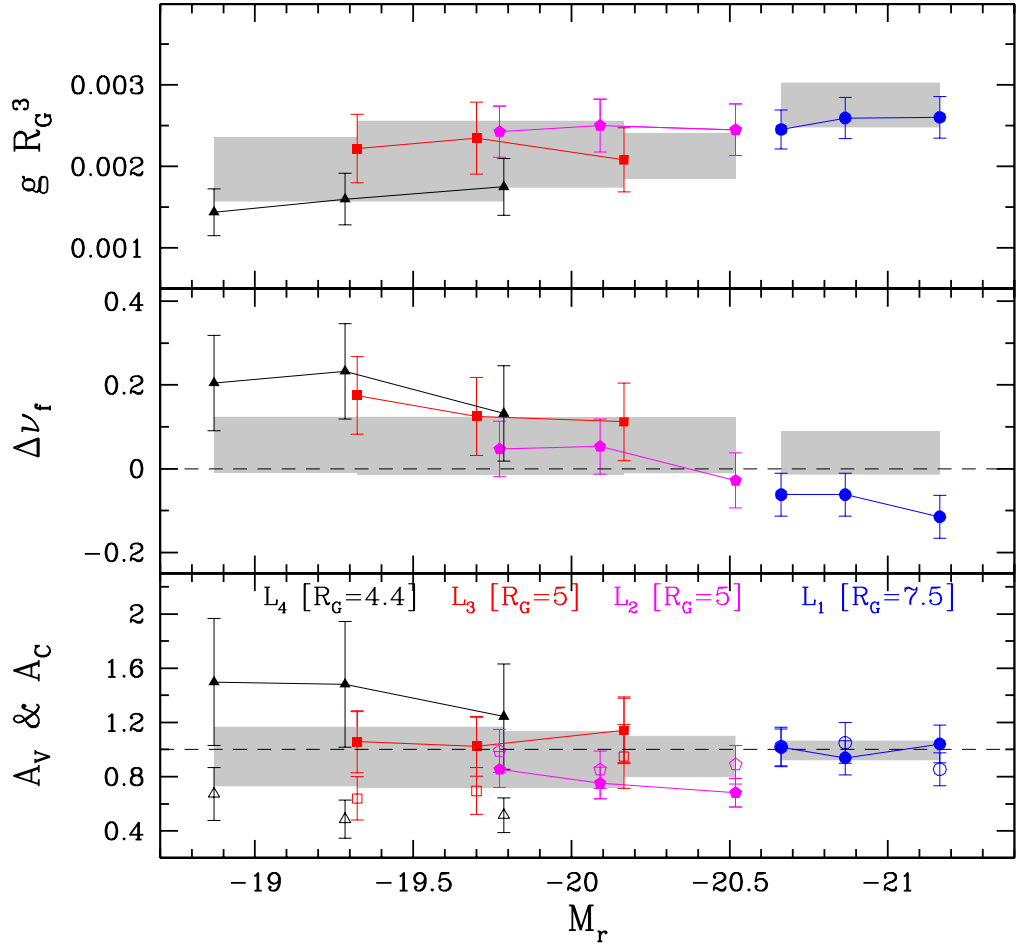}
\caption{Genus-related statistics for luminosity subsamples (see Table~1 for
definitions). The measured values of subsamples that belong to the
same luminosity sample are connected together except for the $A_V$
parameter. The smoothing lengths adopted are $R_G=$ 7.5, 5.0, 5.0, and
4.4 $\hMpc$ for subsamples of L1, L2, L3, and L4, respectively. Shaded
regions are the 1 $\sigma$ variation regions calculated from
100 mock surveys.
In the bottom panel, $A_C$ is given by filled symbols, $A_V$
by open symbols, and the shaded areas are shown only for
$A_C$.
[{\it See the electronic edition of the Journal for a color version
of this figure.}]
}
\label{fig9}
\end{figure}

We find a statistically significant scale dependence in the
amplitude of the genus curve in the case of the Best sample (top panel
of Figure~\ref{fig7}). This scale dependence, which is expected in a
$\Lambda{\rm CDM}$ universe, is consistent with the results of mock
surveys in a $\Lambda{\rm CDM}$ model over the smoothing scales
explored.  

The $A_V$ parameter, a measure of void multiplicity, is found to be
smaller than 1 at all smoothing scales explored, from 3.5 to 11
$\hMpc$. Since this parameter cannot become less than 1 through
gravitational evolution, it provides strong evidence for biased galaxy
formation at low density environments.  Galaxies form in such a way
that under-dense regions can be more connected than expected in the
unbiased galaxy formation process of initially Gaussian fluctuations.
The observed connectivity of voids could also arise for some special
class of initial conditions like the bubbly density field 
in the extended inflationary scenario (La \& Steinhardt 1989).
This measurement using genus statistics provides a new constraint on
models for galaxy biasing.

In our scale dependence study, we note that the shift parameter,
$\Delta\nu$, is more negative and cluster multiplicity parameter,
$A_C$, is smaller than predicted by $\Lambda{\rm CDM}$ for mock
surveys of the Best sample. We attribute these phenomena to the large,
connected high density regions like the Sloan Great Wall and an
{\large $x$}-shaped structure which runs several hundred Mpcs across
the survey volume.  The statistical significance level is about
$3\sigma$.  More realistic mock galaxy catalogs including biasing and
more completed SDSS data are needed to draw conclusions on the
consistency between the $\Lambda{\rm CDM}$ model and the observed
universe. We have also found paucity of galaxies brighter than
about $M_r = -22$ at large distances in the 
Sample 14. Volume-limited samples
constrained by absolute magnitude cuts brighter than $-22$ 
in the Sample 14 have radial
density gradients, and are thus not suitable for measuring parameters
like the genus amplitude.

We clearly detect the signature of luminosity bias of the topology
through the shift parameter $\Delta \nu$.  We find that the genus of
bright galaxies is more negatively (\ie, meatball) shifted than that
of faint ones. The transition from negative to positive shift occurs
at close to the characteristic magnitude $M_{r*}=-20.44$ of SDSS
galaxies. This difference in the topology of bright and faint galaxies
provides a further test for galaxy biasing models.

We test for a scale dependence of this
luminosity bias by comparing the genus-related parameters of the
brighter subset (L2-1) of the luminosity sample L2 with those of the
fainter subset (L2-3) at smoothing lengths 5, 6, and 7
$h^-1$Mpc. Given the uncertainties in the measured statistics, we do
not find such a scale dependence. 
%For example, the shift parameter
%$\Delta\nu$ of the bright subsample L2-1 whose median absolute
%magnitude of galaxies is $-20.52$, is smaller by about 0.1 at all
%three scales than that of the faint sample L2-3 whose median absolute
%magnitude is $-19.77$.
% \msv{If important, these details should go above in the Results.}

\begin{figure}
\epsscale{1.1}
\plotone{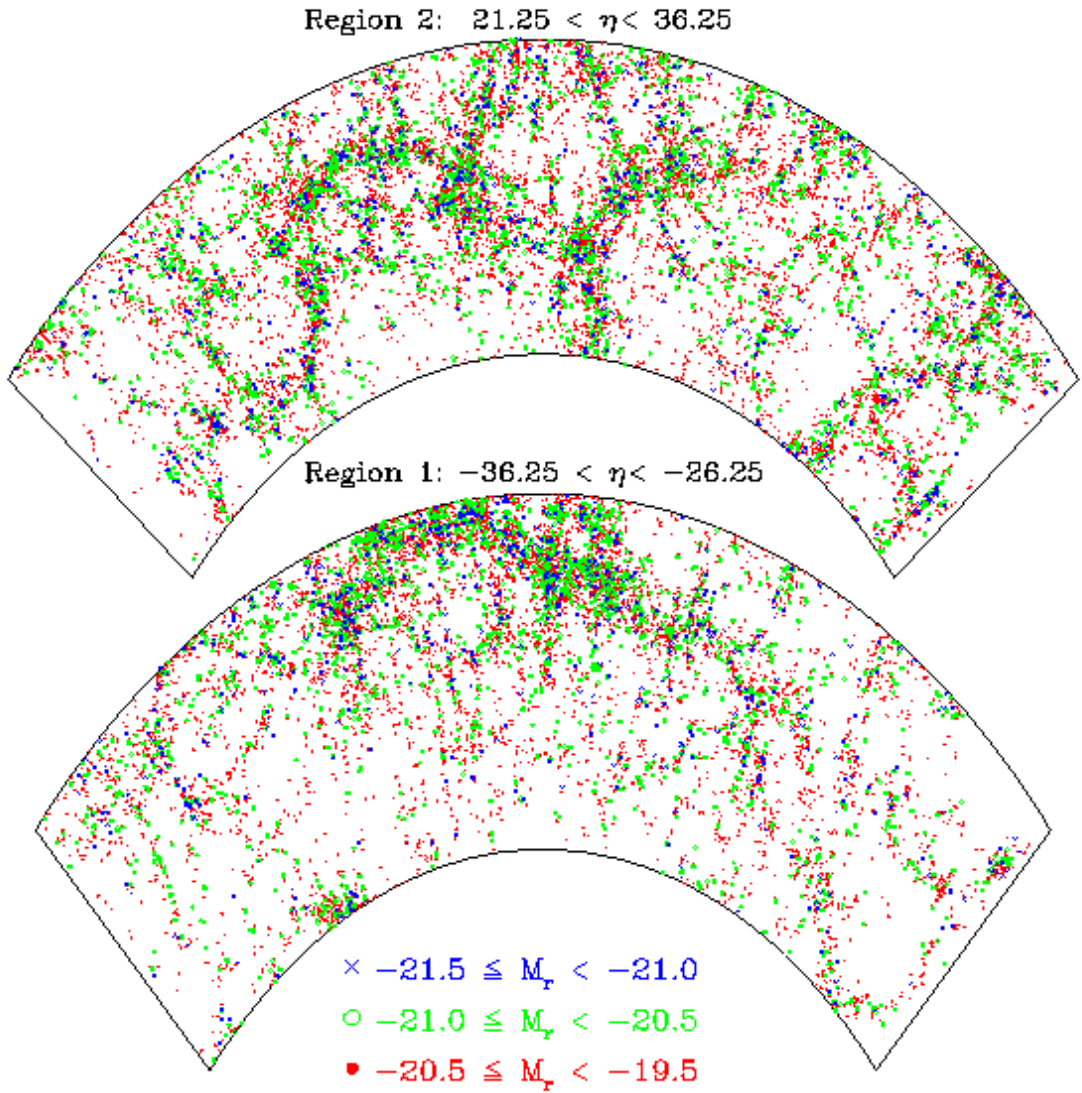}
\caption{Distribution of galaxies in the luminosity sample L2 in the comoving
distance versus survey longitude coordinate plane projected to the median
volume latitude. Galaxies are distinguished by color and point type in
accordance with their absolute magnitudes. Crosses are the brightest,
circles the middle, and dots the faintest.
[{\it See the electronic edition of the Journal for a color version
of this figure.}]
}\label{fig10}
\end{figure}

To visually demonstrate variations in the spatial distribution of
galaxies with different luminosity, in Figure~\ref{fig10} we plot the comoving
distances and survey longitude of galaxies in sample L2.  Crosses
and circles are galaxies brighter than
about $M_{*}$, while dots indicate galaxies fainter than
about $M_*$. It is evident that faint galaxies are found in both low
and high density environments, while brighter galaxies (e.g., crosses
in Figure~10), are rarely found in low density regions.
Because galaxies fainter than $M_{*}$ fill the universe more uniformly,
but are still absent within large tunnels and voids, their
distribution has more of a bubble topology. Galaxies brighter than
$M_{*}$ delineate dense clusters, filaments and walls, surrounded by
large empty spaces which fill most volume of the
universe. Accordingly, they show a shift toward a meatball topology.
The galaxy biasing mechanism should not only make brighter galaxies
hard to form in under-dense environments and cluster more strongly,
but also make the distributions of bright and faint galaxies have
meatball and bubble shifted topologies, respectively. It is not clear
whether or not simple galaxy formation prescriptions like the
semi-analytic and numerical models (Berlind {\etal} 2003; Kravtsov
{\etal} 2004; Zheng {\etal} 2004) or Halo Occupation Distribution
modeling of galaxy formation (Ma \& Fry 2000; Berlind \& Weinberg
2002) satisfy these constraints.  We plan to study this in future
work.

\acknowledgments
CBP and MGP acknowledge the support of the Korea Science and Engineering
Foundation (KOSEF) through the Astrophysical Research Center for the
Structure and Evolution of the Cosmos (ARCSEC) and through the grant
R01-2004-000-10520-0. MSV acknowledges support from NASA grant
NAG-12243. JRG has been supported by NSF grant AST04-06713.

Funding for the creation and distribution of the SDSS Archive has been
provided by the Alfred P. Sloan Foundation, the Participating
Institutions, the National Aeronautics and Space Administration, the
National Science Foundation, the U.S. Department of Energy, the
Japanese Monbukagakusho, and the Max Planck Society. The SDSS Web site
is \url{http://www.sdss.org/}.

The SDSS is managed by the Astrophysical Research Consortium (ARC) for
the Participating Institutions. The Participating Institutions are The
University of Chicago, Fermilab, the Institute for Advanced Study, the
Japan Participation Group, The Johns Hopkins University, the Korean
Scientist Group, Los Alamos National Laboratory, the
Max-Planck-Institute for Astronomy (MPIA), the Max-Planck-Institute
for Astrophysics (MPA), New Mexico State University, University of
Pittsburgh, University of Portsmouth, Princeton University, the United
States Naval Observatory, and the University of Washington.


\clearpage


%%%%%%%%%%%%%%%%%%%% Table 1 %%%%%%%%%%%%%%%%%%%%%%%%%
%\begin{table}[b]\footnotesize
%\begin{center}
%\begin{minipage}{180mm}
%\caption{Volume-Limited Luminosity and Scale Dependence Samples}
%\begin{tabular}{lccccc}
%\hline\hline
%Sample Name &Abs.~Mag & Redshift & Distance $^a$
%& Galaxies  &${N_{\rm res}} $$^b$\\
%\hline
%{\bf Scale Dependence:}\\
\begin{deluxetable}{lccccc}
\tabletypesize{\footnotesize}
\tablecolumns{5}
\tablewidth{39pc}
%\begin{center}
\tablecaption{Volume-Limited Luminosity and Scale Dependence Samples}
\tablehead{
\colhead{Sample Name} &\colhead{Abs.~Mag} & \colhead{Redshift} & \colhead{Distance \tablenotemark{a}}
& \colhead{Galaxies} &\colhead{${N_{\rm res}}\tablenotemark{b}$}
}
\startdata
\sidehead{~~Scale Dependence:}
~~Sparse & $-22.22\leq M_r<-21.20$ & $0.0740<z<0.1654$ & $218.2<r<477.1$& $20138$ & 3618\\
~~Best   & $-21.53\leq M_r<-20.15$ & $0.0550<z<0.1091$ & $162.9<r<319.0$& $36000$ & 6466\\
~~Dense  & $-20.12\leq M_r<-18.44$ & $0.0295<z<0.0529$ & $ 88.0<r<156.8$& $13011$ & 2336\\
\hline
\sidehead{~~Luminosity Dependence:}
%{\bf Luminosity Dependence:}\\
~~ L1& $-22.50\leq M_r<-20.50$ & $0.0833<z<0.1257$ & $245.1<r<366.1$& $27623$ & 4961\\
~~~~{\s L1-1} &{\s $-22.50\leq M_r<-20.87$} &    &&{\s 13811}&{\s 2481} \\
~~~~{\s L1-2} &{\s $-21.16\leq M_r<-20.66$} &    &&{\s 13811}&{\s 2481}\\
~~~~{\s L1-3} &{\s $-20.87\leq M_r<-20.50$} &    &&{\s 13811}&{\s 2481}\\
~~ L2& $-21.50\leq M_r<-19.50$ & $0.0543<z<0.0833$ & $160.9<r<245.1$& $29932$  & 5374 \\
~~~~{\s L2-1} & {\s$-21.50\leq M_r<-20.09$} &    &&{\s 14966}&{\s 2687}\\
~~~~{\s L2-2} & {\s$-20.52\leq M_r<-19.77$} &    &&{\s 14966}&{\s 2687}\\
~~~~{\s L2-3} & {\s$-20.09\leq M_r<-19.50$} &    &&{\s 14966}&{\s 2687}\\
~~ L3& $-21.00\leq M_r<-19.00$ & $0.0436<z<0.0674$ & $129.6<r<198.9$& $19315$  & 3468 \\
~~~~{\s L3-1} & {\s$-21.00\leq M_r<-19.70$} &    &&{\s 9657}&{\s 1734}\\
~~~~{\s L3-2} & {\s$-20.17\leq M_r<-19.32$} &    &&{\s 9657}&{\s 1734}\\
~~~~{\s L3-3} & {\s$-19.70\leq M_r<-19.00$} &    &&{\s 9657}&{\s 1734}\\
~~ L4& $-20.50\leq M_r<-18.50$ & $0.0350<z<0.0543$ & $104.1<r<160.9$& $13107$  & 2353 \\
~~~~{\s L4-1} & {\s$-20.50\leq M_r<-19.29$} &    &&{\s 6553}&{\s 1177}\\
~~~~{\s L4-2} & {\s$-19.76\leq M_r<-18.87$} &    &&{\s 6553}&{\s 1177}\\
~~~~{\s L4-3} & {\s$-19.29\leq M_r<-18.50$} &    &&{\s 6553}&{\s 1177}\\
\enddata
%\end{tabular}
%\end{minipage}
%\end{center}
\tablecomments{The samples
are volume-limited with apparent magnitude limits of $14.5\leq m_r \leq 17.5$.}
\tablenotetext{a}{Comoving distance in units of $h^{-1} \rm Mpc$.}
\tablenotetext{b}{Number of resolution elements
calculated by $\Omega (r_{\rm max}^3-r_{min}^3)/3(2\pi)^{3/2}{\bar d}^3$,
where $\Omega$ is the solid angle of our analysis area and ${\bar d}$ is
the mean separation of galaxies listed in Table 2.}
\end{deluxetable}

\begin{deluxetable}{lcccccc}
\tabletypesize{\footnotesize}
\tablecolumns{7}
%\tablewidth{45pc}
\tablecaption{Genus-Related Statistics of the Observational Samples and Subsamples} 
\tablehead{
\colhead{Sample Name}   &\colhead{$\bar d$\tablenotemark{~a}} &\colhead{$R_{\rm G}$\tablenotemark{~b}}&\colhead{$G_{obs}$}
&\colhead{$\Delta \nu$} &\colhead{$A_{\rm V}$}&\colhead{$A_{\rm C}$}
}
\startdata
\sidehead{~~Scale Dependence:}
~~Sparse& 11.31 & 9.5 & $68.5(56.7){\pm}5.6 $&$ -0.10(-0.02){\pm}0.04 $& $ 0.72(0.80){\pm}0.07 $&$1.29(1.30){\pm}0.12$\\
        &       &11.0 & $41.0(36.0){\pm}4.1 $&$ -0.06(-0.00){\pm}0.05 $& $ 0.74(0.80){\pm}0.09 $&$1.45(1.45){\pm}0.17$\\
~~Best  & 6.14  & 5.0 &$125.4(103.3){\pm}9.7$&$ -0.17(-0.05){\pm}0.03 $& $ 0.69(0.78){\pm}0.05 $&$0.80(0.86){\pm}0.08$\\
        &       & 6.0 & $77.3(69.7){\pm}7.3 $&$ -0.15(-0.08){\pm}0.05 $& $ 0.71(0.75){\pm}0.07 $&$0.74(0.75){\pm}0.08$\\
        &       & 7.0 & $52.3(49.8){\pm}5.6 $&$ -0.15(-0.10){\pm}0.06 $& $ 0.70(0.72){\pm}0.09 $&$0.74(0.73){\pm}0.10$\\
        &       & 8.0 & $36.2(35.2){\pm}4.9 $&$ -0.15(-0.11){\pm}0.07 $& $ 0.74(0.75){\pm}0.12 $&$0.69(0.66){\pm}0.11$\\
        &       & 9.0 & $25.0(24.8){\pm}4.0 $&$ -0.145(-0.12){\pm}0.08$& $ 0.79(0.79){\pm}0.16 $&$0.74(0.71){\pm}0.14$\\
~~Dense & 4.17  & 3.5 & $24.1(19.7){\pm}4.1 $&$  0.18(0.32){\pm}0.078 $& $ 0.42(0.46){\pm}0.08 $&$1.21(1.34){\pm}0.28$\\
        &       & 4.0 & $15.6(13.6){\pm}3.2 $&$  0.18(0.27){\pm}0.094 $& $ 0.54(0.57){\pm}0.13 $&$1.26(1.30){\pm}0.34$\\
        &       & 4.5 & $10.4(9.4) {\pm}2.4 $&$  0.19(0.26){\pm}0.106 $& $ 0.46(0.46){\pm}0.14 $&$1.14(1.17){\pm}0.37$\\
\tableline
\sidehead{~~Luminosity Dependence:}
~~L1          &    7.17  &    6.0 &    $ 85.3(71.8)\pm7.5$  &    $-0.11(-0.01)\pm0.04$ &    $0.80(0.89)\pm0.08$ &    $ 0.98(1.03)\pm0.11$ \\
~~~~{\s L1-1} &{\s 9.04} &{\s 7.5}&{\s $ 47.2(39.3)\pm4.6$} &{\s $-0.14(-0.12)\pm0.05$}&{\s $0.76(0.85)\pm0.11$}&{\s $ 1.01(1.04)\pm0.13$}\\
~~~~{\s L1-2} &{\s 9.04} &{\s 7.5}&{\s $ 47.0(39.2)\pm4.6$} &{\s $-0.09(-0.06)\pm0.05$}&{\s $0.93(1.05)\pm0.13$}&{\s $ 0.91(0.94)\pm0.12$}\\
~~~~{\s L1-3} &{\s 9.04} &{\s 7.5}&{\s $ 44.5(37.1)\pm4.3$} &{\s $-0.09(-0.06)\pm0.05$}&{\s $0.91(1.02)\pm0.13$}&{\s $ 0.98(1.02)\pm0.13$}\\
~~L2     &         4.71  &    4.0 &    $ 80.8(67.4)\pm8.6$   &   $-0.04( 0.07)\pm0.05$ &    $0.81(0.89)\pm0.10$ &    $ 0.87(0.95)\pm0.11$ \\
~~~~{\s L2-1} &{\s 5.94} &{\s 5.0}&{\s $ 46.9(39.2)\pm6.1$} &{\s $-0.13(-0.03)\pm0.07$}&{\s $0.80(0.89)\pm0.12$}&{\s $ 0.65(0.68)\pm0.10$}\\
~~~~{\s L2-2} &{\s 5.94} &{\s 5.0}&{\s $ 47.9(40.0)\pm6.2$} &{\s $-0.05(0.05) \pm0.07$}&{\s $0.77(0.85)\pm0.12$}&{\s $ 0.72(0.75)\pm0.11$}\\
~~~~{\s L2-3} &{\s 5.94} &{\s 5.0}&{\s $ 46.4(38.8)\pm6.0$} &{\s $-0.06(0.05) \pm0.07$}&{\s $0.90(0.99)\pm0.14$}&{\s $ 0.82(0.85)\pm0.13$}\\
~~L3          &    4.44  &    3.8 &    $ 39.1(33.3)\pm6.0$   &   $ 0.04(0.16) \pm0.07$ &    $0.62(0.67)\pm0.09$ &    $ 1.13(1.21)\pm0.20$ \\
~~~~{\s L3-1} &{\s 5.54} &{\s 5.0}&{\s $ 19.5(16.5)\pm3.7$} &{\s $ 0.02(0.11) \pm0.09$}&{\s $0.89(0.95)\pm0.20$}&{\s $ 1.11(1.14)\pm0.24$}\\
~~~~{\s L3-2} &{\s 5.54} &{\s 5.0}&{\s $ 22.0(18.6)\pm4.1$} &{\s $ 0.03(0.13) \pm0.09$}&{\s $0.65(0.69)\pm0.15$}&{\s $ 0.99(1.02)\pm0.22$}\\
~~~~{\s L3-3} &{\s 5.54} &{\s 5.0}&{\s $ 20.8(17.6)\pm3.9$} &{\s $ 0.08(0.18) \pm0.09$}&{\s $0.60(0.64)\pm0.14$}&{\s $ 1.03(1.06)\pm0.22$}\\
~~L4          &    4.09  &    3.5 &    $ 23.3(19.4)\pm3.9$  &    $ 0.12(0.24) \pm0.09$ &    $0.44(0.48)\pm0.09$ &    $ 1.23(1.34)\pm0.31$ \\
~~~~{\s L4-1} &{\s 5.16} &{\s 4.4}&{\s $ 12.8(10.5)\pm2.5$} &{\s $ 0.02(0.13) \pm0.11$}&{\s $0.47(0.52)\pm0.12$}&{\s $ 1.15(1.24)\pm0.36$}\\
~~~~{\s L4-2} &{\s 5.16} &{\s 4.4}&{\s $ 11.7(9.6) \pm2.3$} &{\s $ 0.12(0.23) \pm0.11$}&{\s $0.44(0.49)\pm0.12$}&{\s $ 1.37(1.48)\pm0.43$}\\
~~~~{\s L4-3} &{\s 5.16} &{\s 4.4}&{\s $ 10.5(8.7) \pm2.1$} &{\s $ 0.09(0.21) \pm0.11$}&{\s $0.61(0.67)\pm0.16$}&{\s $ 1.38(1.50)\pm0.43$}\\
\enddata
\tablecomments{$G_{obs}$ is the amplitude of the
observed genus curve, $\Delta \nu$ is shift parameter, and
$A_C$ and $A_V$ are cluster and void abundance parameters, respectively.
Uncertainty limits are estimated from
mock surveys in redshift space, and systematic
bias-corrected values are given in parentheses.}
\tablenotetext{a}{Mean separation in units of $h^{-1}\rm Mpc$.}
\tablenotetext{b}{Smoothing length in units of $\hMpc$.}
\end{deluxetable}
%%%%%%%%%%%%%%%%%%%%%% END of TABLE 2 %%%%%%%%%%%%%%%%%%%%%%%

\end{document}